\documentclass[12pt]{iopart}

\usepackage{iopams}  

\expandafter\let\csname equation*\endcsname\relax
\expandafter\let\csname endequation*\endcsname\relax

\usepackage{amsmath}
\usepackage{graphicx}

\usepackage{appendix}

\usepackage{color}
\newcommand{\new}{\textcolor{blue}}

\newcommand{\kav}{{\bra k \ket}}
\newcommand{\bra}{{\langle}}
\newcommand{\ket}{{\rangle}}
\newcommand{\bA}{{\bf A}}
\newcommand{\bk}{{\bf k}}
\newcommand{\btauh}{{\boldsymbol{ \hat{\tau}}}}
\newcommand{\hr}{{\hat{r}}}

\newcommand{\bq}{{{\bf q}}}
\newcommand{\Brh}{{{\boldsymbol{\hat{ r}}}}}
\newcommand{\hatg}{{\hat{g}}}
\newcommand{\bzet}{{\boldsymbol{\zeta}}}

\begin{document}
	
	\title[Vaccination with partial transmission and social distancing]{Vaccination with partial transmission and social distancing on contact networks}

	\author{Christian John Hurry\textsuperscript{1}, Alexander Mozeika\textsuperscript{2}, Alessia Annibale\textsuperscript{1,3}}
	
	\address{\textsuperscript{1}Department of Mathematics, King's College London, Strand, WC2R 2LS \\
		\textsuperscript{2}School of Cancer and Pharmaceutical Sciences, King’s College London, London, UK\\
		\textsuperscript{3}Institute for Mathematical \& Molecular Biomedicine, Hodgkin Building, Guy's Campus, London, SE1 1UL}
	\ead{christian.hurry@kcl.ac.uk \\
		alexander.mozeika@kcl.ac.uk\\
		alessia.annibale@kcl.ac.uk}
	\vspace{10pt}
	\begin{indented}
		\item[]Date: Oct 2021
	\end{indented}
	
	\begin{abstract}
		We study the impact of vaccination on the risk of epidemics spreading through structured networks using the cavity method of statistical physics. We relax the assumption that vaccination prevents all transmission of a disease used in previous studies, such that vaccinated nodes have a small probability of transmission. To do so we extend the cavity method to study networks where nodes have heterogeneous transmissibility.  We find that vaccination with partial transmission still provides herd immunity and show how the herd immunity threshold depends upon the assortativity between nodes of different transmissibility. In addition, we study the impact of social distancing via bond percolation and show that percolation targeting links between nodes of high transmissibility can reduce the risk of an epidemic greater than targeting links between nodes of high degree.
		Finally, we extend recent methods to compute the distributional equations of risk in populations with heterogeneous transmissibility and show how targeted social distancing measures may reduce overall risk greater than untargeted vaccination campaigns, by comparing the effect of random and targeted strategies of node and link deletion on the risk distribution.
	\end{abstract}
	
	%
	\vspace{2pc}
	\submitto{\JSTAT}
	
	%
	%

	\maketitle
	\newpage

	\section{Introduction}
	
	The study of compartmental models, where a population is split into, for example, `Susceptible', `Infected' and `Recovered' compartments in the SIR model, have formed an important theoretical and computational basis for the study of epidemics and design of vaccination campaigns. An increased understanding of contact networks, which detail physical contacts of a sustained duration between individuals in a population (see e.g. \cite{Salathe2010}), has been complemented by mathematical results for compartmental epidemic models on networks. One such contribution is the prediction of an epidemic threshold dependent upon the rates of infection and recovery, as well as the contact network topology \cite{DallAsta2005,Newman2002a,Gleeson2008}. Many important results concerning the SIR model on networks have been derived by message-passing approaches, also known as the cavity method, including the size and risk of
	epidemics on networks with arbitrary degree distributions \cite{Karrer2010} and degree correlations \cite{Vazquez2003, Shiraki2010}. More recently, these methods have been employed to model epidemic mitigation via contact tracing apps \cite{bianconi2021message} and competing strains of infectious diseases \cite{Min2020,Sun2021,Moore2020}. In addition to analytical results, the cavity method provides a set of equations that allows for an efficient \emph{parallel} numerical implementation, as opposed to direct simulations of the SIR model, which are usually \emph{sequential} and typically marred by long computation times, scaling with the size and connectivity of the network and the infection rate relative to the recovery rate.
	
	One of the key problems that mathematical epidemiology addressed is the optimisation of vaccination strategies, i.e given a finite supply of vaccines, who should be vaccinated to mitigate an epidemic. Previous work has studied vaccination strategies based upon network topology.  Vaccination which prioritises nodes of higher degree, i.e. people with a high number of social contacts, is known to lead to better outcomes in comparison with a random vaccination campaign \cite{Pastor-Satorras2002a}. Furthermore, it has been shown that vaccination using information beyond the degree can improve upon degree-based strategies \cite{Miller2007,Ma2013}. The recent discovery of the role that the Hashimoto non-backtracking matrix \cite{Hamilton2014,Karrer2014,Rogers2015} plays in epidemic processes has led to the use of eigenvalues to rank nodes for prioritisation of vaccination \cite{torres2021nonbacktracking}. The benefit of modelling such strategies is that they provide principles upon which vaccination campaigns can be based with only partial knowledge of the contact network. For example, it can provide theoretical insight into prioritising vaccination for parts of the population with higher than average social connectivity. 
	
	Vaccination is usually modelled by placing an individual into a separate vaccinated compartment, such that vaccinated nodes block incoming infections, so that if enough people are vaccinated the infection can no longer spread through the population, a phenomena referred to as herd immunity. This assumes that a vaccine provides full protection against transmission of an infectious disease for any individual. In general, it may be desirable to relax this assumption. For example, although the efficacy of a vaccine in preventing symptoms is ascertained before approval for public use, it is more difficult to determine how a given vaccine prevents transmission until data is collected during or after vaccine roll-out. In this case, it may be desirable to assume that vaccination reduces the transmissibility of an individual to a small but finite value. Furthermore, it is also important to recognise that vaccines are usually given to priority groups first, such as people with underlying health conditions, medical staff, or people above a certain age. Due to the correlated nature of social contacts \cite{Mossong2008,Mistry2021}, prioritising vaccinations may lead to correlations between the vaccinated status of a node and its topological properties. Variation in the transmissibility of individuals has been explored in previous works \cite{Newman2002a,ANDERSON1984,Gou2017,baxter2021degree}, however, analytical results have been restricted to the average risk or size of the epidemic. It has previously been shown that there is high variability in the risk of individual nodes, due to differences in the node environment \cite{Rogers2015,Kuhn2017}, which is neglected when only considering the average risk. In our work, we show that this variability is more prominent when transmissibility varies between individuals and focus on how this impacts the distribution of risk.
	
	After reviewing node percolation approaches to vaccination, we focus our study on vaccination with partial transmission by extending the cavity approach to allow for \emph{heterogeneity} in the transmission between individuals. At first we consider transmissibility and degree to be uncorrelated, and we derive the herd immunity threshold in this scenario, showing that vaccination with partial transmission will always require a greater proportion of the population vaccinated to achieve herd immunity. We then relax the assumption that transmissibility and degree are uncorrelated to show how these correlations affect epidemic risk. 
	Using this framework, we then study the impact of social distancing between groups of different transmissibility, by deriving equations for the risk of an epidemic under link percolation that targets either links between nodes of high transmissibility, or nodes of high degree. To go beyond the average risk of an epidemic, we follow techniques developed in context of the micro-structure of percolation in the configuration model \cite{Kuhn2017} to investigate the distribution of risk in structured and heterogeneous populations.
	Our analysis reveals a highly non-trivial distribution of risk, 
	even amongst nodes of the same degree and transmissability. 
	Finally, we extend the cavity approach for the distributional equations of risk to account for node and link deletion. We comment that this provides a succinct procedure to explore the impact of targeted vaccination and social distancing strategies on the distribution of risk.

	The remaining sections of this article are organised as follows. In Sec. \ref{sec: perfect vacc} we review the network percolation approach to the study of the steady state reached in the SIR model on 
	networks with arbitrary degree distributions and degree correlations.
	In Sec. \ref{sec: heterogeneous transmission} we extend the cavity method to account for a network with nodes assigned to sub-types of different transmissibility, and provide a closed set of equations for the average risk a node poses to the network. We consider the cases where node transmissibility is and is not correlated with social contact, separately. In Sec. \ref{sec: social distancing} we derive equations for the average risk under link percolation, and show how the choice of the links which are deleted affects the risk. Sec. \ref{sec: dist of risk} demonstrates how the cavity method may be used to derive the distributional equations of risk and how to solve these equations via a population dynamics procedure. We conclude with a discussion of the theoretical value of our results and potential avenues for future work. 
	Technical details are described in the appendices.

	\section{Impact of vaccination on the epidemic risk in contact networks} \label{sec: perfect vacc}
	One of the main interests in the study of the risk of epidemics spreading on 
	contact networks is the exploration of vaccination strategies. A question that arises, 
	when assessing vaccination strategies, 
	is what fraction of the population needs to be vaccinated to achieve herd immunity. In this section we 
	consider the SIR epidemic model on contact networks and we show that this question can be answered by studying a simple node 
	percolation problem.
	
	\subsection{Herd immunity with perfect vaccination in the SIR model on structured networks}
	We consider the SIR model on an undirected network with $N$ nodes and adjacency matrix $\bA$. The elements $A_{ij}=A_{ji} \in \{0,1\}~\forall~i,j$ denote presence ($A_{ij}=1$) or absence ($A_{ij}=0$) of link for each pair of nodes $(i,j)$ and 
	$A_{ii}=0~\forall~i$. 
	Following the formulation in \cite{Rogers2015}, we 
	assume that an infected node may pass an infection to a neighbouring node in a short time interval $\rmd t$ with probability $\beta\rmd t$ and an infected node recovers from an infection at a time drawn from an `infectious time' distribution, $\gamma(t)$, normalised such that $\int_{0}^{\infty} \rmd t \gamma(t) = 1$, to ensure that an infected node eventually recovers. In order to quantify the fraction of the population that needs to be vaccinated to achieve herd immunity, we note 
	that if a vaccine provides perfect immunity, its action 
	on an individual is functionally equivalent to 
	deleting a node in the network, as this prevents the spread of infection
	through that node and its links. In order to incorporate the effect of node deletion we  
	define a binary variable $\sigma_{i} \in \{0,1\}$ which describes whether a node has been vaccinated ($\sigma_{i}=0$) or not ($\sigma_{i}=1$). 
	As detailed in \cite{Rogers2015}, the probability $r_i$ that an infection starting from a single node $i$ causes an outbreak across the bulk of the network, also called the \emph{risk} of node $i$, can be determined, via a message-passing approach. Using the same approach and incorporating the effect of node deletion, we can write
	\begin{eqnarray}
		r_{i}(\bA) &=\sigma_{i}\left[1 - \int_{0}^{\infty}\rmd t\,\gamma(t)\prod_{j \in \partial^\bA_{i}}\left(1 - \left(1 - \rme^{-\beta t}\right) r_{j}^{(i)}(\bA)\right)\right] \label{eq:r} \\ 
		r_{j}^{(i)}(\bA) &= \sigma_{j}\left[ 1 - \int_{0}^{\infty}\rmd t\,\gamma(t)\prod_{\ell \in \partial^\bA_{j}\setminus i}\left(1 - \left(1 - \rme^{-\beta t}\right) r_{\ell}^{(j)}(\bA)\right)\right] \label{eq:rhat}
	\end{eqnarray}
	where $\partial^\bA_i=\{j:A_{ij}=1\}$ is the set of the neighbours of $i$ and 
	$r_{j}^{(i)}(\bA)$ is the risk that node $j$ causes an outbreak in the `cavity' graph where node $i$ is removed. These equations are exact on trees, where loops are 
	absent, and become exact on locally tree-like graphs, in the limit of large system size, where the typical length of loops diverges logarithmically \cite{Melnik2011}. 
	For a given contact network, one can solve the cavity equations (\ref{eq:r}) and (\ref{eq:rhat})
	numerically. 
	Alternatively, in absence of true knowledge of the 
	contact network one can assume that it is \emph{random} and use equations (\ref{eq:r}) and (\ref{eq:rhat}) to 
	derive equations for the global risk 
	$g(\bA)= \frac{1}{N}\sum_{i=1}^N r_{i}(\bA)$, which is expected to be \emph{self-averaging} \cite{mezard1987spin} when $N\rightarrow\infty$ and thus independent of its microscopic details. 
	Averaging (\ref{eq:r})
	over all sites as shown in \ref{app:homo}, one obtains, for a single graph instance $\bA$
	\begin{eqnarray}
		\hspace*{-1.5cm}g(\bA) &\hspace*{-0.15cm}=& \sum_\sigma \sigma\left[\mathrm{P}(\sigma|\bA) - \sum_{k}\mathrm{P}(k,\sigma|\bA)\int_{0}^{\infty}\rmd t\, \gamma(t) \left(1 - \alpha(t)\hat{g}_{k}(\bA)\right)^{k}\right] \label{eq:g-sigma-A}\\
		\hspace*{-1.6cm}\hat{g}_{k}(\bA) &\hspace*{-0.2cm}=& \sum_{\sigma'}\sigma'\left[\mathrm{P}(\sigma'|\bA) - \sum_{k'\geq 1}\frac{\mathrm{W}(k; k',\sigma'|\bA)}{\mathrm{W}(k|\bA)}\int_{0}^{\infty}\rmd t\, \gamma(t) \left(1 - \alpha(t)\hat{g}_{k'}(\bA)\right)^{k'-1}\right]\nonumber
		\label{eq:ghat-sigma}
	\end{eqnarray}
	where we have denoted for brevity $1-\rme^{-\beta t}=\alpha(t)$. We have defined the likelihood to draw at random a node with degree $k$ and label $\sigma$ for the graph instance $\bA$ as  $\mathrm{P}(k,\sigma|\bA)=N^{-1}\sum_i\delta_{\sigma,\sigma_i(\bA)}\delta_{k,k_i(\bA)}$, with $k_i(\bA)=|\partial^\bA_i|$ denoting the degree of node $i$ and $\mathrm{P}(\sigma|\bA)=\sum_k \mathrm{P}(k,\sigma|\bA)$ the marginal distribution. Furthermore, we have defined $\mathrm{W}(k;k',\sigma'|\bA)=\sum_{ij}A_{ij}\delta_{k,k_i(\bA)}\delta_{k',k_j(\bA)}\delta_{\sigma',\sigma_j(\bA)}/N\bar{k}(\bA)$ as the likelihood that by drawing a link at random we choose a link with a node of degree $k$ at one end and a node with degree $k'$ and label $\sigma'$ at the other. We denote by
	$\bar k(\bA)=N^{-1}\sum_i k_i(\bA)$ the mean degree.  The distribution $\mathrm{W}(k;k'|\bA) = \sum_{\sigma}\mathrm{W}(k;k',\sigma|\bA)$ is known as the degree correlations, and 
	$\mathrm{W}(k|\bA)=\sum_{k'}\mathrm{W}(k;k'|\bA)$ is its marginal distribution. 
	Here and below we adopt the convention to denote a joint probability distribution of node quantities across connected node pairs with $\mathrm{W}(\dots;\dots)$. We will consider vaccination strategies based upon the degree of nodes, such that $\mathrm{P}(k,\sigma|\bA)=\mathrm{P}(k|\bA)\mathrm{P}(\sigma|k)$ and $\mathrm{W}(k;k',\sigma'|\bA)=\mathrm{W}(k;k'|\bA)\mathrm{P}(\sigma'|k')$, where $\mathrm{P}(\sigma|k)$ models the degree-dependent vaccination strategy. 
	
	In the limit of large networks, $N\to\infty$, we expect
	fluctuations of intensive network observables to
	vanish and their value on a single graph realisation to coincide with their ensemble averaged value. Assuming 
	that the degree distribution and degree correlations
	display such self-averaging behavior in the limit of large $N$, we average the cavity equations  
	over a suitably defined random graph ensemble, obtaining 
	\begin{eqnarray}
		g &= \sum_\sigma \sigma\left[\mathrm{P}(\sigma) - \sum_{k}\mathrm{P}(k,\sigma)\int_{0}^{\infty}\rmd t\, \gamma(t) \left(1 - \alpha(t)\hat{g}_{k}\right)^{k}\right] \label{eq:g-sigma}\\
		\hat{g}_{k} &= \sum_{k'\geq 1} \sum_{\sigma'}\mathrm{W}(k'|k) \mathrm{P}(\sigma'|k')\sigma'\left[1 - \int_{0}^{\infty}\rmd t\, \gamma(t) \left(1 - \alpha(t)\hat{g}_{k'}\right)^{k'-1}\right] \label{eq:ghat-W}
	\end{eqnarray}
	where we have introduced the ensemble averages $\mathrm{P}(\sigma)=\bra \mathrm{P}(\sigma|\bA)\ket_\bA$, $\mathrm{P}(k,\sigma)=\bra \mathrm{P}(k,\sigma|\bA)\ket_\bA$, $\mathrm{W}(k'|k')=\bra \mathrm{W}(k'|k,\bA)\ket_\bA$ and 
	$\bra k \ket=\bra \bar k(\bA)\ket_\bA$, with $\bra \cdot \ket_\bA=\sum_\bA \cdot \mathrm{P}(\bA)$ and $\mathrm{P}(\bA)$ the probability over the set of symmetric adjacency matrices $\bA\in\{0,1\}^{N(N-1)/2}$, which defines the random graph ensemble. 
	
	The set of equations (\ref{eq:ghat-W}) may be solved numerically. It is easy to check that they always have a trivial solution 
	$\hat{g}_{k}=0\, \forall ~k$, corresponding to the absence of an epidemic 
	outbreak, i.e. $g=0$. 
	The stability of this solution depends upon the Jacobian of the system of equations (\ref{eq:ghat-W}), such that an epidemic will occur if $|\lambda_{1}^{M}| > 1$ where $\lambda_{1}^{M}$ is the largest eigenvalue of the matrix with entries
	\begin{align}
		M_{k,k'} = (k'-1)\mathrm{W}(k'|k)\sum_{\sigma'}\sigma'\mathrm{P}(\sigma'|k')T \label{eq: homo degree corr matrix outbreak cond}\new{,}
	\end{align}
	where $T = \int_{0}^{\infty}\rmd t\, \gamma(t)\alpha(t)$ is known as the transmissability of 
	the nodes \cite{Karrer2010}. The above result was first 
	derived in \cite{Vazquez2003}, in absence of node deletion. 
	For uncorrelated networks, $\mathrm{W}(k'|k)=\mathrm{W}(k')$ and one has  $\hat{g}_k=\hat{g}~\forall ~k\geq 1$ with 
	\begin{eqnarray}
		\hspace{-2cm}\hat{g} &= \sum_{\sigma} \sigma \left[\mathrm{P}(\sigma) - \sum_{k\geq 1}  \mathrm{P}(\sigma|k)\frac{k\mathrm{P}(k)}{\kav}\int_{0}^{\infty}\rmd t\, \gamma(t) \left(1 - \alpha(t)\hat{g}\right)^{k-1}\right]\nonumber \equiv f(\hat{g})
		\label{eq:ghat-unc}
	\end{eqnarray}
	Due to the monotonic (i.e. $f'(\hat{g})>0$) and concave 
	(i.e. $f''(\hat{g})<0$) nature of $f(\hat{g})$, 
	a non-trivial solution 
	exists, and is stable, if $f'(0)>1$. Evaluating this gives a condition for an epidemic to occur, 
	\begin{align}
		T > \frac{\kav}{\sum_{k,\sigma} \sigma \mathrm{P}(\sigma|k)  k(k-1) \mathrm{P}(k)}.
	\end{align}
	Focusing on the simplest case of a random vaccination campaign, where the variables  
	$\{\sigma_i\}_{i=1}^N$ do not depend on the network degrees and are random i.i.d. with distribution
	\begin{eqnarray}
		\mathrm{P}(\sigma) = p\delta_{\sigma,1} + (1-p)\delta_{\sigma,0} \label{eq:p-sigma} ,
	\end{eqnarray}
	where $(1-p)\in [0,1]$ describes the fraction of nodes that are vaccinated, one finds a critical value of $p$ below which epidemics are prevented,
	\begin{eqnarray}
		p<p_c=\frac{\kav}{T\bra k(k-1)\ket}.
		\label{eq:thresh-unc-p}
	\end{eqnarray}
	This is the herd immunity threshold, a rearranged form of the percolation threshold in the configuration model 
	\cite{Newman2018}.

	\begin{figure}[t]
		\includegraphics[width=0.49\textwidth]{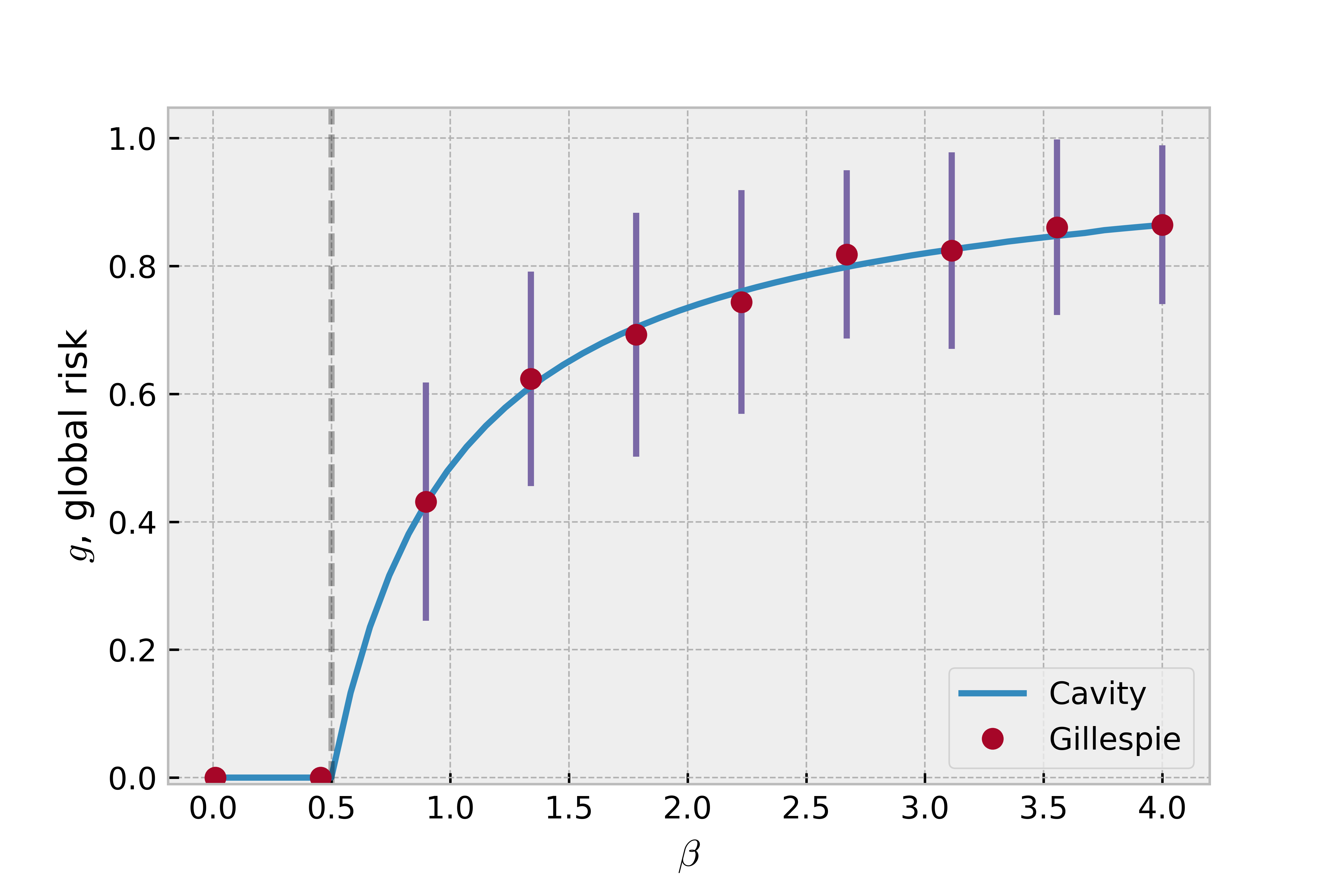} \includegraphics[width=0.49\textwidth]{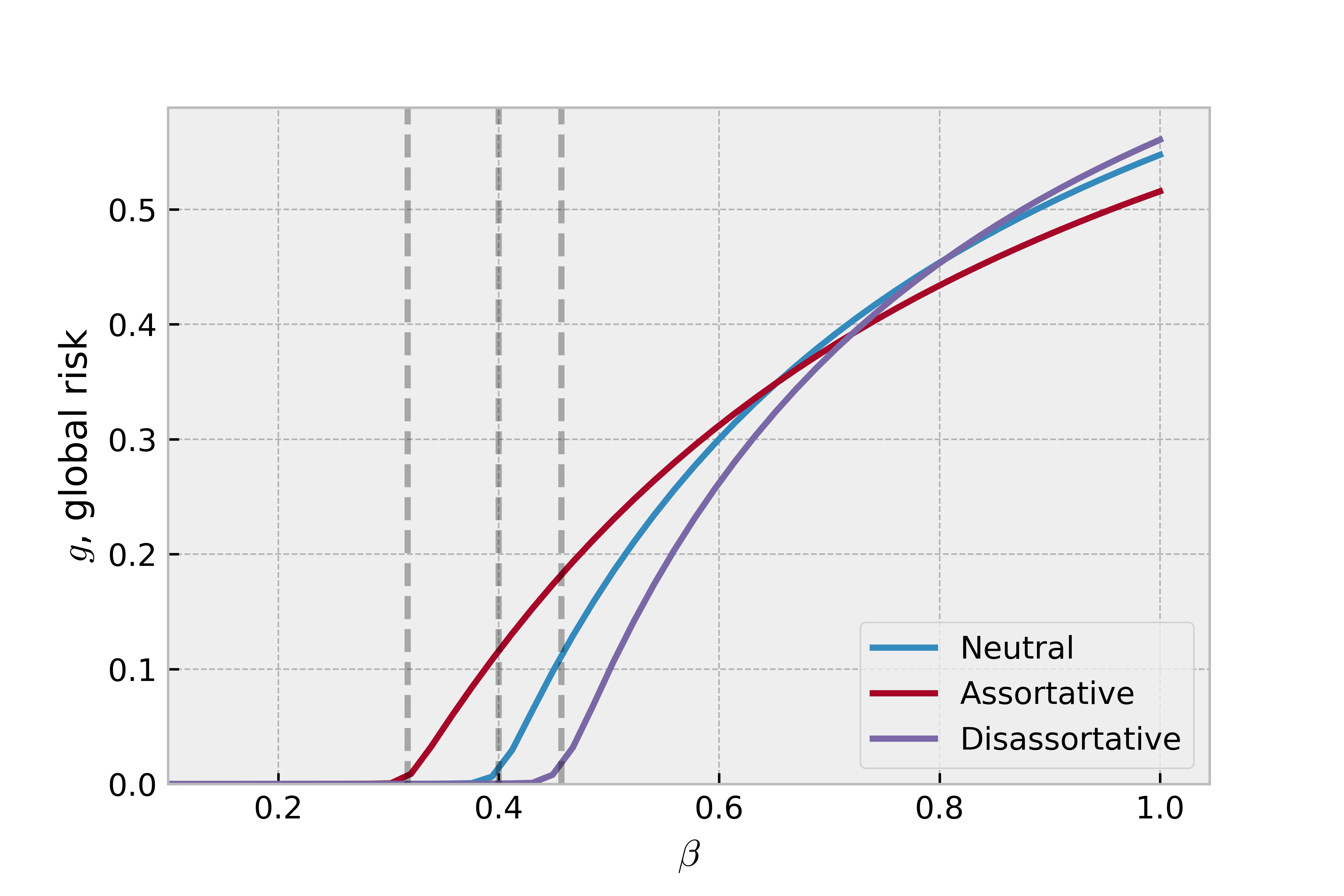}
		\caption{ Global risk as a function of the infection rate $\beta$, for nodes with exponential infectious time distribution.
			Left: Results shown for an Erd\"os-R\'enyi (ER) graph with average connectivity $\kav=5$ and nodes with mean infectious time $\xi^{-1}=1/2$. Solid lines indicate theoretical predictions from the cavity equations (\ref{eq:ghat-W}) while symbols show results from simulations on an ER graph of size $N=3000$, averaged over $100$ initial sites of infection, and $20$ runs starting from each site.  The dashed line indicating the epidemic threshold is computed from equation (\ref{eq:thresh-unc-p}), which simplifies to $p_c=1/T\kav$ for ER graphs. Error bars show the standard deviation of the risk across different initial sites of infection. Right: Theoretical predictions from equation (\ref{eq:ghat-W}) are shown for  graphs with Poissonian degree distribution  with $\kav=3$ and nodes with mean infectious time $\xi^{-1}=1.25$. Results are shown for graphs with neutral, assortative and disassortative degree correlations. Dashed lines indicate the predicted epidemic threshold from the largest eigenvalue of (\ref{eq: homo degree corr matrix outbreak cond}) }
		\label{fig: risk against beta}
	\end{figure}

\begin{figure}[t!]
		\includegraphics[width=0.49\textwidth]{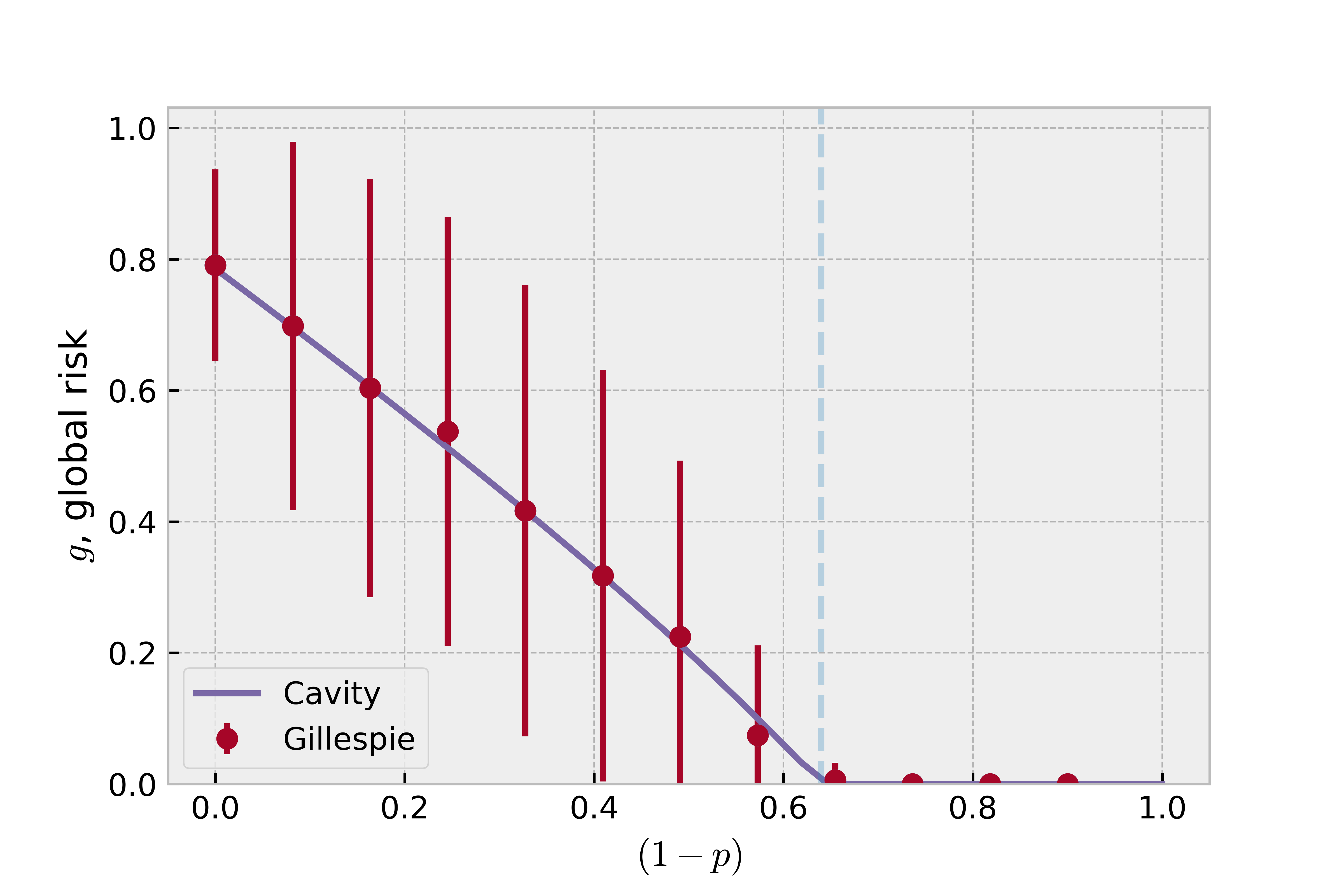} 
		\includegraphics[width=0.49\textwidth]{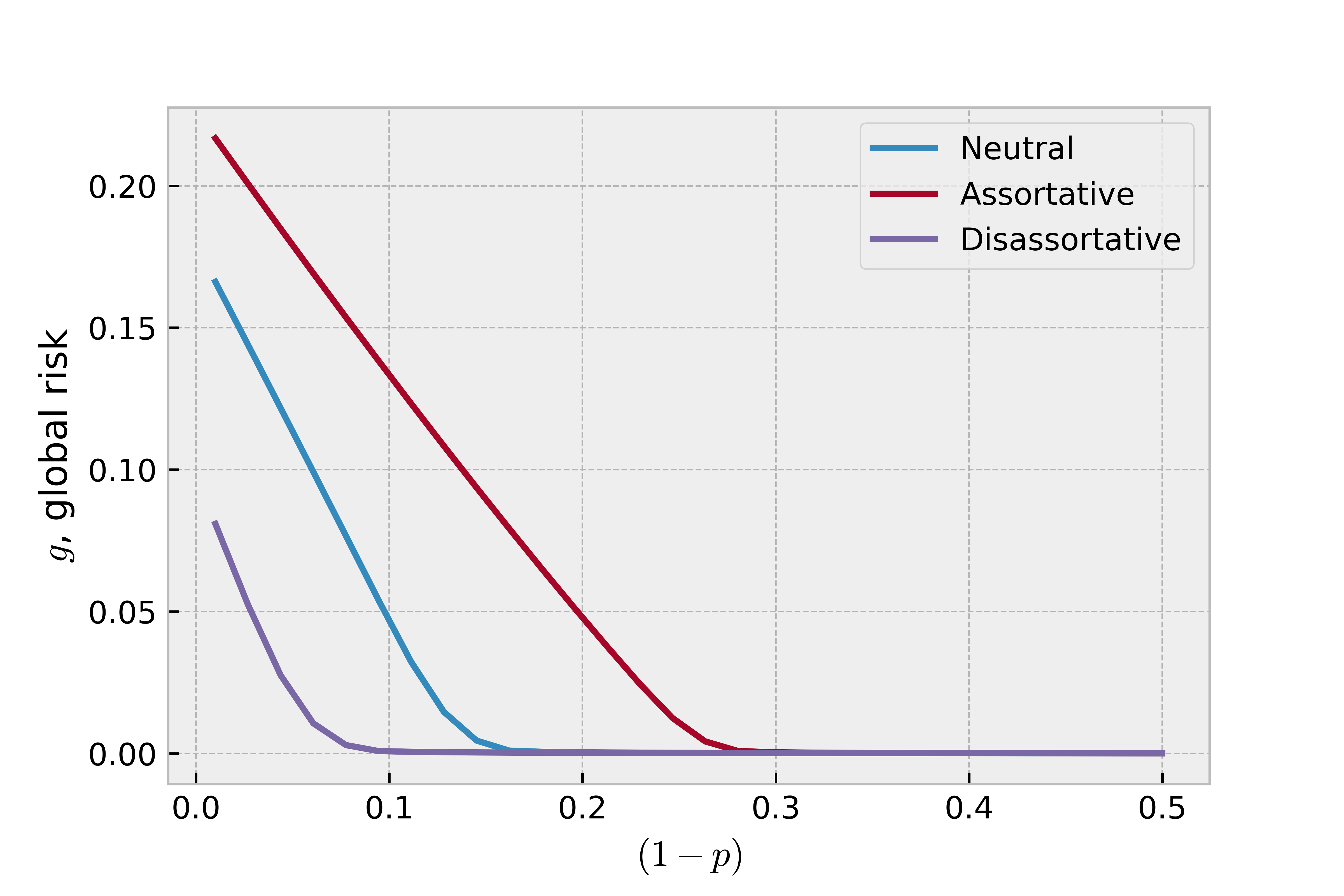} 
		\caption{Global risk plotted as a function of the fraction $1-p$ of vaccinated nodes with infection rate $\beta=0.5$ and exponential distribution of infectious times with parameter $\xi=0.4$. Left: For an Erd\"os-R\'enyi graph with $\kav=5$ we show predictions (solid line) from the cavity equations (\ref{eq:ghat-W}) and results of simulations (circles)
			on graphs with size $N=1000$, where nodes were selected with probability $1-p$ to be vaccinated, 
			averaged over $10$ configurations of vaccinated nodes, $25$ initial sites of infection, and $25$ runs. The dashed line indicates the herd immunity as computed from equation (\ref{eq:thresh-unc-p}). Right: Theoretical results from the cavity equations (\ref{eq:ghat-W}) are shown for graphs with Poissonian degree distribution, mean degree $\kav=3$, for neutral, assortative and disassortative degree correlations}
		\label{fig: homo vacc }
	\end{figure}

	In Figure \ref{fig: risk against beta} (left panel) the global risk predicted from the cavity equation (\ref{eq:g-sigma}) is plotted for an Erd\"os-R\'enyi (ER) graph with mean degree $\kav=5$ and shown to be in excellent agreement with results from simulations \cite{Gillespie1977,kiss2017mathematics,Miller2019} of the stochastic SIR model on networks with $N=3000$ nodes.  
	We note that the wide error bars, showing the standard deviation of the risk across different initial sites of infection, indicate that there is large variation in the risk of individual nodes in a given ER network.  The impact of degree correlations is shown on the right panel of Figure \ref{fig: risk against beta}, where results from the cavity equations are 
	shown for networks with the same average connectivity and Poissonian degree distribution, but assortative and disassortative degree mixing, respectively. To compute the degree correlation function $\mathrm{W}(k,k')$ for assortative and disassortative Poissonian graphs we followed the methods described in \cite{Annibale2009}. These results are consistent with results for the size of the giant component in networks with degree correlations \cite{Newman2002}. In Figure \ref{fig: homo vacc } we plot the global risk as a function of the fraction $1-p$ of vaccinated population in ER graphs with average connectivity $\kav=5$. In the left panel we see that predictions from the cavity equations (\ref{eq:ghat-W}) are in agreement with results from simulations on networks of size $N=1000$. The right panel shows that the herd immunity threshold is lowest in Poissonian graphs with disassortative correlations, and highest for ER graphs with assortative correlations.

	\section{Impact of vaccination with partial transmission}\label{sec: heterogeneous transmission}
	In the previous section we have considered an idealised 
	vaccine that prevented all infection passing through paths containing vaccinated nodes.
	In general, vaccinations may not prevent all transmission, and individuals may respond to vaccines differently, leading to heterogeneities in individual infectious potential.
	It is not possible to know \emph{a priori} how an individual will respond to a vaccine, however it is feasible to gather data on the impact of vaccination on transmissibility within a population and estimate differences in transmission between different demographics. It is therefore of interest to understand how variations in coverage and infectious potential of individuals can affect vaccination strategies. In order to do so, we extend the cavity approach to the SIR model to account for heterogeneity in the transmissibility of individuals. We then go on to show how this affects the herd immunity threshold.

	\subsection{Epidemic risk with heterogeneous transmissibility}
	
	In order to account for heterogeneity in the 
	transmissibility of individuals, 
	we assume that individuals remain infectious for different 
	times. 
	Without loss of generality we assume that the infectious
	time distribution of each individual has the same functional form but its scale varies between individuals of different sub-type.
	Therefore, we introduce $M$ node sub-types, identified by labels $\xi \in \{\xi^{1},...,\xi^{M}\}$, and assign each node a label, $\{\xi_i\}_{i=1}^N$, according to some probability distribution $\mathrm{P}(\xi)$. The labels $\{\xi^{1},...,\xi^{M}\}$ take values in $(0,\infty)$, 
	such that $\xi$ parameterises the infectious time distribution $\gamma(t|\xi)$ of nodes with label $\xi$.
	For the remainder of this paper we choose $\gamma(t|\xi) = \xi \rme^{-\xi t}$ such that $\frac{1}{\xi}$ represents the \emph{mean infectious time} of individuals with label $\xi$. This particular choice for the recovery time distribution is known as Markovian recovery, 
	however, our approach also holds for non-Markovian recovery times, which could be implemented by e.g. a Weibull distribution \cite{Kiss2015}.
	
	In this case, the risk of node $i$ depends on the label of node $i$, and all labels downstream of node $i$. We denote the set of labels downstream of node $i$, including node $i$, by $\boldsymbol{\xi}_{i}$. 
	The equations for the local risk for a given 
	instance of a contact network $\bA$ with prescribed labels $\bxi=(\xi_1,\ldots,\xi_N)$ read as
	\begin{eqnarray}
		r_{i}(\bA,\boldsymbol{\xi}_{i}) &= 1 - \int_{0}^{\infty}\rmd t\,\gamma(t|\xi_{i})\prod_{j \in \partial_{i}^\bA}\left(1 - \alpha (t) r_{j}^{(i)}(\bA,\boldsymbol{\xi}_{j})\right) \label{eq:r-hetero} \\ 
		r_{j}^{(i)}(\bA,\boldsymbol{\xi}_{j}) &= 1 - \int_{0}^{\infty}\rmd t\,\gamma(t|\xi_{j})\prod_{\ell \in \partial_{j}^\bA\setminus i}\left(1 - \alpha (t) r_{\ell}^{(j)}(\bA,\boldsymbol{\xi}_{\ell})\right).
		\label{eq:rhat-hetero}
	\end{eqnarray}
	Averaging over the nodes, we find the equation for the global 
	risk, which depends on the whole label sequence $\bxi$ and the network $\bA$
	\begin{eqnarray}
		g(\bA,\bxi)=   \frac{1}{N} \sum_{i=1}^{N} r_{i}(\bA,\bxi).
	\end{eqnarray}
	In \ref{app:hetero} we show that the assumption of a random tree-like structure allows us to find a closed expression 
	for the global risk in the $N\to \infty$ limit, that once averaged over the graph and label ensemble, reads as 
	\begin{align}
		g &= 1 - \sum_{k,\xi} \mathrm{P}(k,\xi) \int_{0}^{\infty} \rmd t \gamma(t|\xi) \left(1 - \alpha(t) \hat{g}_{k ,\xi}\right)^{k} \label{eq:hetero-global-risk} \\
		\hat{g}_{k, \xi}  &= \sum_{k ',\xi'} \mathrm{W}(k',\xi'|k,\xi)\left[1 - \int_{0}^{\infty} \rmd t \gamma(t|\xi') \left(1 - \alpha(t) \hat{g}_{k', \xi'}\right)^{k'-1}\right] \label{eq:hetero-ghat}
	\end{align}
	where $\mathrm{P}(k,\xi)$ and $\mathrm{W}(k',\xi'|k,\xi)$ are the ensemble-averaged values of $\mathrm{P}(k,\xi|\bA,\bxi)=\frac{1}{N}\sum_i \delta_{k,k_i(\bA)}\delta_{\xi,\xi_i}$ and $\mathrm{W}(k',\xi'|k,\xi,\bA,\bxi)= \mathrm{W}(k,\xi;k',\xi'|\bA,\bxi)/ \mathrm{W}(k,\xi|\bA,\bxi)$, respectively, where $\mathrm{W}(k,\xi ;k',\xi'|\bA,\bxi) =\frac{1}{N\bar{k}(\bA)}\sum_{ij} A_{ij}\delta_{k,k_i(\bA)}\delta_{k',k_j(\bA)}\delta_{\xi,\xi_i}\delta_{\xi',\xi_j}$. For a given kernel $\mathrm{W}(k',\xi'|k,\xi)$ of a network with maximum degree $k_{\max}$ one may solve the system of $M \times k_{\max}$ equations (\ref{eq:hetero-ghat}) numerically and substitute the result into (\ref{eq:hetero-global-risk}) to find the global risk. 

	Alternatively, one may be able to simplify these equations for specific choices of $\mathrm{W}(k,\xi;k',\xi')$. For example, if we assume the transmissibility of an individual is uncorrelated with its degree, the degree of its neighbours and the transmissibility of its neighbour, in this case $\mathrm{W}(k,\xi;k',\xi')=\mathrm{W}(k;k')\mathrm{P}(\xi)\mathrm{P}(\xi')$, and upon substitution of this kernel into the r.h.s of (\ref{eq:hetero-ghat}) we see that $\hatg _{k,\xi} = \hatg_{k} \forall \xi$, and the system of equations (\ref{eq:hetero-ghat}) reduce to, 
	\begin{align}
		\hat{g}_{k} &= \sum_{k'} \mathrm{W}(k'|k) \left[1 -  \int_{0}^{\infty} \rmd t\left<\gamma(t|\xi)\right>_{\xi} \left(1 - \alpha(t) \hat{g}_{k'}\right)^{k'-1} \right] \label{eq:hetero-ghat-label-uncorr}
	\end{align}
	where $\left<...\right>_{\xi}=\sum_{\xi} ... \mathrm{P}(\xi)$. We see that equation (\ref{eq:hetero-ghat-label-uncorr}) takes the same form as (\ref{eq:ghat-W}) but now depends upon the `average' infectious time distribution $\left<\gamma(t|\xi)\right>_{\xi}$. The stability criteria is therefore identical to that of (\ref{eq:ghat-W}) but now depends upon the average transmissibility, such that an epidemic will occur if
	\begin{align}
		|\lambda_{1}^{(\bf{J})} |> 1
	\end{align}
	where $\lambda_{1}^{(\bf{J})}$ is the largest eigenvalue of the Jacobian of (\ref{eq:hetero-ghat-label-uncorr}) with entries $J_{k,k'} = \left<T(\xi)\right>_{\xi} (k'-1)\mathrm{W}(k'|k)$ and $T(\xi) = \int_{0}^{\infty} \rmd t\gamma(t|\xi) \alpha(t)$. A further simplification can be made if we assume the degrees are uncorrelated, in which case equations (\ref{eq:hetero-ghat-label-uncorr}) reduces to a single equation, 
	\begin{align}
		\hat{g} &= 1 - \sum_{k} \frac{k\mathrm{P}(k)}{\kav} \int_{0}^{\infty} \rmd t\left<\gamma(t|\xi)\right>_{\xi} \left(1 - \alpha(t) \hat{g}\right)^{k-1} \label{eq:hetero-ghat-all-uncorr}
	\end{align}
	which allows a non-zero solution when 
	the epidemic threshold exceeds a value dependent on the average transmissibility, 
	\begin{eqnarray}
		\frac{\bra k^2\ket-\bra k \ket}{\kav}\geq\frac{1}{\left<T(\xi)\right>_{\xi}}. \label{eq:epithresh-hetero-alluncorr}
	\end{eqnarray}

	Another interesting case is where the degrees of nodes at either end of a randomly selected link are conditionally uncorrelated given their labels $\xi$ and $\xi'$ i.e when $\mathrm{W}(k,\xi;k',\xi') = \mathrm{W}(\xi,\xi')\mathrm{W}(k|\xi)\mathrm{W}(k'|\xi')$. For this choice of $\mathrm{W}(k,\xi;k',\xi')$ we find that $\mathrm{W}(k',\xi'|k,\xi) = \frac{\mathrm{W}(k,\xi;k',\xi')}{  \mathrm{W}(k,\xi)} =  \frac{\mathrm{W}(\xi;\xi')\mathrm{W}(k|\xi)\mathrm{W}(k'|\xi')}{\mathrm{W}(k|\xi)\mathrm{W}(\xi)}$. As a result of this, the cavity equations (\ref{eq:hetero-ghat}) simplify such that $\hatg_{k, \xi} = \hatg_{\xi} ~\forall ~k $ where, 
	\begin{eqnarray}
		\hatg_{\xi} &= 1 - \sum_{k',\xi'}\mathrm{W}(\xi'|\xi) \mathrm{W}(k'|\xi') \int_{0}^{\infty} \rmd t\gamma(t|\xi') \left(1 - \alpha(t) \hat{g}_{\xi'}\right)^{k'-1}. \label{eq:hetero-ghat-ER}
	\end{eqnarray}
	By using this form of $\mathrm{W}(k,\xi;k',\xi')$ we have reduced (\ref{eq:hetero-ghat}) from a system of $k_{\max} \times M$ equations to just $M$ equations. This is a significant reduction, particularly for networks with large maximum degree $k_{\max}$. We note that although we have chosen $\xi$ to label nodes of different transmissibility, we could choose $\gamma(t|\xi) = \gamma(t)$ and allow $\xi$ to label some other node property (e.g age). Equations (\ref{eq:hetero-ghat-ER}) would therefore provide a quick numerical implementation to find epidemic risk in networks where degree correlations are generated by a node property, $\xi$, according to $\mathrm{W}(k,\xi;k',\xi') = \mathrm{W}(\xi,\xi')\mathrm{W}(k|\xi)\mathrm{W}(k'|\xi')$.
	By analysis of the Jacobian of the system of equations (\ref{eq:hetero-ghat-ER}), we find that an epidemic will occur if $|\lambda_{1}^{(J_{\xi,\xi'})}|>1$ where $\lambda_{1}^{(J_{\xi,\xi'})}$ is the largest eigenvalue of $J_{\xi,\xi'} = \sum_{k'}\mathrm{W}(\xi'|\xi) (k'-1) \mathrm{W}(k'|\xi')T(\xi')$. 
	
	\subsection{Herd immunity for vaccines with partial transmission }
	
	Having extended the cavity approach to account for heterogeneity in individual transmissibility, we can now consider vaccination with partial transmission. 
	In this case, vaccinated individuals can still catch and transmit the disease, but we assume their transmissability is reduced, such that an individual $i$ who is vaccinated will have a lower mean infectious time $1/\xi_i$ than an unvaccinated individual. To analyse this scenario, we split a population into two clusters: an unvaccinated population with low infectious time decay rate $\xi^{\ell}$ and a vaccinated population with high decay rate $\xi^{h} > \xi^{\ell}$. The fraction of the population that is vaccinated is given by $\mathrm{P}(\xi^h) = 1 - \mathrm{P}(\xi^\ell)$. To keep the analysis simple, we focus on random vaccinations, 
	where one can assume that the labels $\xi$ are not correlated with the network degrees, and we will assume 
	that degrees are uncorrelated. 
	The critical fraction 
	$P_c(\xi^h)$ of population that needs to be vaccinated to prevent epidemics, assuming vaccination with partial transmission, can be found from (\ref{eq:epithresh-hetero-alluncorr}) by writing 
	\begin{eqnarray}
		\bra T(\xi)\ket_\xi&=&\mathrm{P}(\xi^h)T(\xi^h)+[1-\mathrm{P}(\xi^h)]T(\xi^\ell)
		\nonumber\\
		&=&T(\xi^\ell)-\mathrm{P}(\xi^h)\Delta 
	\end{eqnarray}
	with $\Delta=T(\xi^\ell)-T(\xi^h)>0$, which, substituted 
	in  
	(\ref{eq:epithresh-hetero-alluncorr}), gives
	\begin{eqnarray}
		\mathrm{P}(\xi^h)>\frac{T(\xi^\ell)}{T(\xi^\ell)-T(\xi^h)}\left[1-\frac{\kav}{ T(\xi^\ell)\left<k(k-1)\right>} \right]
		\equiv P_c(\xi^h)
		\label{eq:thresh-xi-hilo}
	\end{eqnarray}
	Clearly, the fraction of the population that has to be vaccinated to achieve herd immunity must be greater when the vaccination leads to partial transmission, as opposed to no transmission. The herd immunity threshold under vaccination without transmission is given by $1-p_{c}$ from (\ref{eq:thresh-unc-p}), where $T$ is the transmissibility of the unvaccinated population i.e $T = T(\xi^{\ell})$.  One can express the herd immunity threshold under vaccination with partial transmission, $P_c(\xi^h)$, in terms of $1 - p_{c}$,
	\begin{eqnarray}
		P_c(\xi^h)=\frac{T(\xi^\ell)}{T(\xi^\ell)-T(\xi^h)}(1-p_c).
	\end{eqnarray}
	Since $T(\xi^\ell)>T(\xi^\ell)-T(\xi^h)$, it is clear that vaccination with partial transmission requires a greater number of people to be vaccinated when compared with vaccination with no transmission i.e $P_c(\xi^h) > 1-p_{c}$. 
	
	\subsection{Networks with correlated structure and transmissibility}
	In general, social contacts will be correlated with the transmissibility of individuals. For example, vaccinations are often prioritised for at risk groups, and for people above a certain age. Furthermore, vaccine distribution is dependent upon supply chains, and this can lead to higher levels of vaccination in one geographic area to another. Aside from vaccinations, transmissibility may correlate with social contacts, for example differences in transmissibility between children and adults have been noted for some infectious diseases, and hence if a social interaction network shows correlation with age, this may affect the spread of epidemics.
	
	To understand how this affects the epidemic risk and vaccination we need to specify an ensemble of networks where links between individuals are based upon their respective transmissibility. 
	For simplicity we do not specify any hard constraints on the degree sequence and instead define an ensemble of networks with average degree $\kav$ where
	links are drawn in a way that 
	allows for preferential attachment between nodes with specific transmissibilities, on the basis of an arbitrary 
	function $\mathrm{W}(\xi;\xi')$ of the labels of the two nodes concerned. More specifically, we assume that 
	for every pair of nodes $i$ and $j$, links are drawn 
	randomly and independently with probability 
	\begin{eqnarray}
		\mathrm{P}(A_{ij}) =  \frac{\kav}{N}\frac{\mathrm{W}(\xi_{i};\xi_{j})}{ \mathrm{P}(\xi_{i})\mathrm{P}(\xi_{j})}\delta_{A_{ij},1}+\left(1-\frac{\kav}{N}\frac{\mathrm{W}(\xi_{i};\xi_{j})}{ \mathrm{P}(\xi_{i})\mathrm{P}(\xi_{j})}
		\right)\delta_{A_{ij},0}
		\label{eq: ModER ensemble}
	\end{eqnarray}
	Here, $\mathrm{W}(\xi;\xi')$ can be interpreted as the probability to draw a link with node labels $\xi$ and 
	$\xi'$ at either end,  
	hence it must be non-negative and normalised to one $\sum_{\xi,\xi'}\mathrm{W}(\xi;\xi')=1$. Due to the undirected nature of the links, we also have $\mathrm{W}(\xi;\xi')=\mathrm{W}(\xi';\xi)$.
	We will refer to nodes with the 
	same label, $\xi$, as a \emph{cluster}, 
	hence $\mathrm{W}$ controls 
	the links between clusters. 
	In \ref{sec: app degree dist} we show that the degree distribution of a graph drawn from this ensemble is given by
	\begin{eqnarray}
		\mathrm{P}(k) =  \sum_{\xi} \mathrm{P}(\xi)\mathrm{P}(k|\xi) =\sum_{\xi} \mathrm{P}(\xi) \rme^{-\bar{k}(\xi)}\frac{(\bar{k}(\xi))^{k}}{k!}
	\end{eqnarray}
	where the average degree is $\kav=\sum_\xi \mathrm{P}(\xi)\bar{k}(\xi)$. This is such that each cluster of nodes in this ensemble has a Poissonian degree distribution $\mathrm{P}(k|\xi)$ with mean degree $\bar{k}(\xi)$, where $\xi$ is the cluster label. Furthermore, we find that the degree correlations are of the special form 
	\begin{eqnarray}
		\mathrm{W}(k,\xi;k',\xi') = \mathrm{W}(\xi;\xi')\mathrm{W}(k|\xi)\mathrm{W}(k'|\xi'),
		\label{eq:corr-k-xi}
	\end{eqnarray}
	where $\mathrm{W}(k|\xi) = k \mathrm{P}(k|\xi)/\bar{k}(\xi)$, which allows us to use the reduced form of the cavity equations (\ref{eq:hetero-ghat-ER}) to study the risk on graphs from this ensemble. Finally, we note that the marginal of $\mathrm{W}(\xi;\xi')$ is $\mathrm{W}(\xi)=\sum_{\xi'}\mathrm{W}(\xi;\xi')=\mathrm{P}(\xi)\bar{k}(\xi)/\kav$.

	\begin{figure}[t]
		\includegraphics[width=0.49\textwidth]{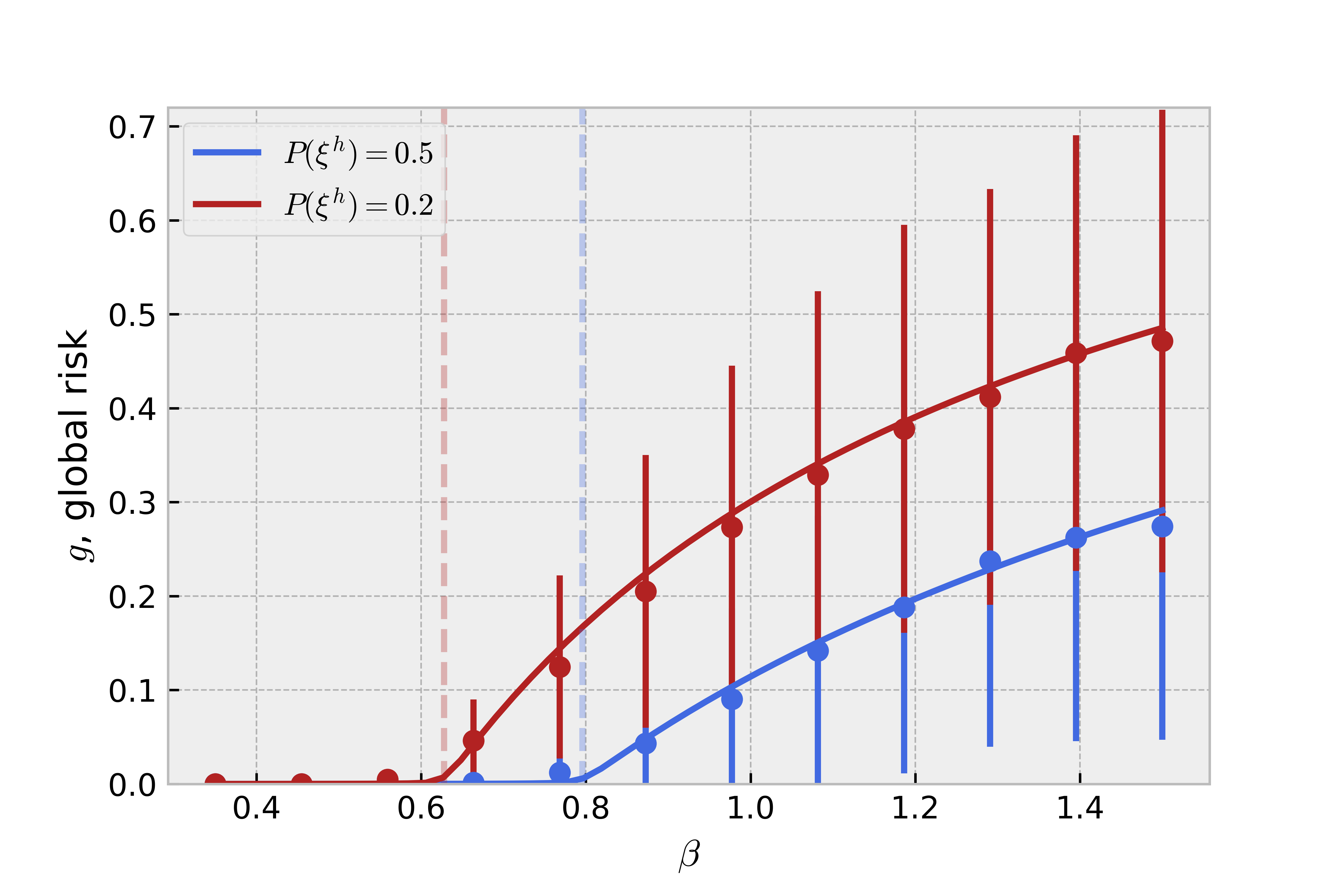}
		\includegraphics[width=0.49\textwidth]{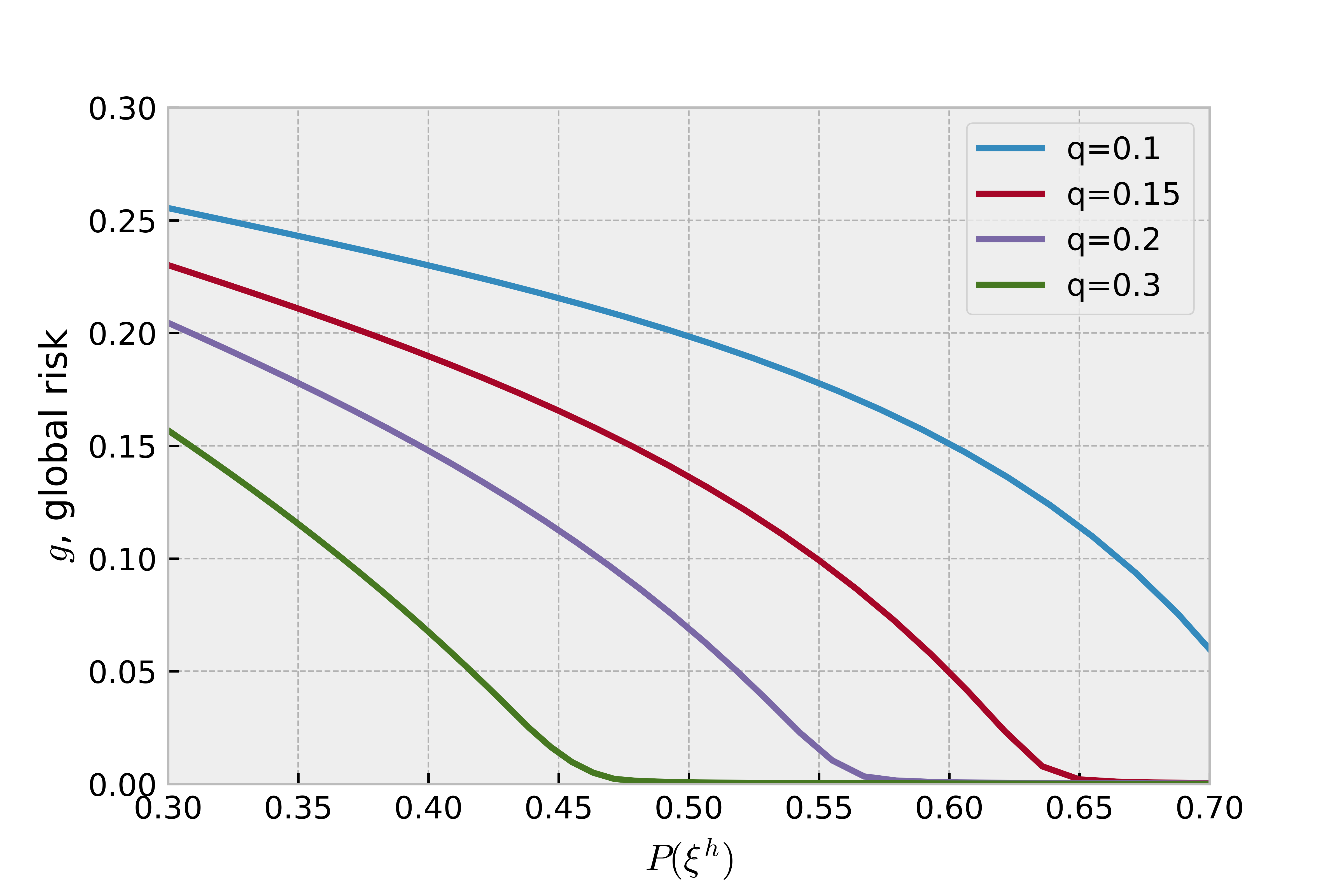}
		\caption{Left: Global risk plotted against infection rate $\beta$ for a population where nodes have an infectious time distribution $\gamma(t|\xi) = \xi \rme^{-\xi t}$ and are separated into two sub-types with $\xi \in\{\xi^{\ell},\xi^{h}\}= \{1.0,5.0\}$. The network is drawn from the ensemble (\ref{eq: ModER ensemble}) with $\kav=3$ and 
			the kernel $\mathrm{W}$ defined in (\ref{eq: two cluster kernel}) with 
			$q=0.15$ and $\bar{k}(\xi^\ell)=\kav$ (and $\bar{k}(\xi^h)=\kav$, via (\ref{eq:kav_kbar})). Results from cavity equations (lines) and simulations (symbols) are shown. Simulations were performed on networks of size $N=2000$, averaged over 500 different sites of initial infection and 100 repetitions. Dashed lines indicate epidemic threshold predicted by the Jacobian of (\ref{eq:hetero-ghat-ER}). Right: For the same population, we plot the global risk against the fraction of fast recovery nodes $\mathrm{P}(\xi^{h})$. We set the infection rate to $\beta=1$. Results from the cavity equations (\ref{eq:hetero-ghat-ER}) are shown for different inter-connectivity $q=\{0.1,0.15,0.2,0.3\}$ (curves from top to bottom). }
		\label{fig: two cluster global risk}
	\end{figure}

	In the remainder of this section, we demonstrate results for networks with two clusters, labelled by  
	$\xi^h$ and $\xi^\ell < \xi^h$, 
	so that the transmissability kernel
	$\mathrm{W}(\xi;\xi')$ is a $2\times 2$ matrix, $\mathrm{P}(\xi^h)=1-\mathrm{P}(\xi^\ell)$ and the average connectivity is related to the intra-cluster connectivities by \begin{eqnarray}
		\kav&=&\bar{k}(\xi^\ell)+\mathrm{P}(\xi^h)(\bar{k}(\xi^h)-\bar{k}(\xi^\ell))
		\nonumber\\
		&=&\bar{k}(\xi^h)+\mathrm{P}(\xi^\ell)(\bar{k}(\xi^\ell)-\bar{k}(\xi^h)).
		\label{eq:kav_kbar}
	\end{eqnarray}
	Given the properties of symmetry, marginalization and normalization of $\mathrm{W}$, we can parameterise 
	$\mathrm{W}(\xi,\xi')$ in terms of 
	two control parameters, 
	the intra-cluster connectivity $\bar{k}(\xi^\ell)$ and 
	a parameter, $q$, that controls the inter-cluster connectivity, 
	via the matrix,
	\begin{equation}
		\mathrm{W} = 
		\begin{pmatrix}
			\frac{\bar{k}(\xi^\ell)\mathrm{P}(\xi^\ell)}{\kav}-q & q \\
			q & 1-\frac{\bar{k}(\xi^\ell)\mathrm{P}(\xi^\ell)}{\kav}-q
		\end{pmatrix} \label{eq: two cluster kernel}
	\end{equation}
	Since the elements of $\mathrm{W}$ are probabilities, they must take values in $[0,1]$, hence the free parameters can take 
	values in the range $q\kav/\mathrm{P}(\xi^\ell)<\bar{k}(\xi^\ell)<(1-q)\kav/\mathrm{P}(\xi^\ell)$ with $q\in [0,1/2]$. 
	
	In Figure \ref{fig: two cluster global risk}  (left panel) the global risk, as predicted from the cavity equations, is plotted for a population with fast and slow recovery nodes, as a function of the infection rate $\beta$, for different values of the fast recovery population density, $\mathrm{P}(\xi^{h})$, and it is found in good agreement with simulations. Figure \ref{fig: two cluster global risk} (right panel) shows that as the proportion of nodes of fast recovery increases, herd immunity is eventually reached. Interestingly, 
	as the inter-connectivity $q$ is increased, the global risk decreases, as does the herd immunity threshold. This can intuitively be explained by first considering the scenario where $q=0$ and the network is split into two disconnected sets of nodes, one where an epidemic is unlikely to be caused by any node, the other where nodes have a non-zero probability of causing an epidemic. As 
	$q$ is increased, the number of links between these 
	sets of nodes increases, with the fast recovery nodes acting as a blockade to the path of infection, lowering the global risk and herd immunity threshold. 
	
	\section{Social distancing in populations with heterogeneous transmissibility}\label{sec: social distancing}
	
	We now show how social distancing affects epidemic risk in a population with heterogeneous transmissibility. We model social distancing as a bond percolation process. We introduce the random binary variables $\tau_{ij} \in \{0,1\}$ which indicates whether a link between two individuals is broken due to social distancing ($\tau_{ij}=0$) or not ($\tau_{ij}=1$). These are drawn from the distribution, 
	\begin{eqnarray}
		\mathrm{Q}(\btau| \mathbf{y}) =\prod_{i<j}\left[ (1 - y_{ij})\delta_{\tau_{ij},0} + y_{ij} \delta_{\tau_{ij},1}\right]
	\end{eqnarray}
	where $1 - y_{ij}$ is the probability that a link between $i$ and $j$ is broken. Under bond percolation the equations for risk take the form, 
	\begin{eqnarray}
		r_{i}(\bA,\bxi) &= 1 - \int_{0}^{\infty}\rmd t\,\gamma(t|\xi_{i})\prod_{j \in \partial^\bA_{i}}\left(1 - \tau_{ij}\alpha(t) r_{j}^{(i)}(\bA,\bxi)\right) \label{eq: r_i hetero link percolation}\\ 
		r_{j}^{(i)}(\bA,\bxi) &= 1 - \int_{0}^{\infty}\rmd t\,\gamma(t|\xi_{j})\prod_{\ell \in \partial^\bA_{j}\setminus i}\left(1 - \tau_{jl}\alpha(t) r_{\ell}^{(j)}(\bA,\bxi)\right). \label{eq: rj_i hetero link percolation}
	\end{eqnarray}
	
\begin{figure}[t]
		\centering
		\includegraphics[width=0.52\textwidth]{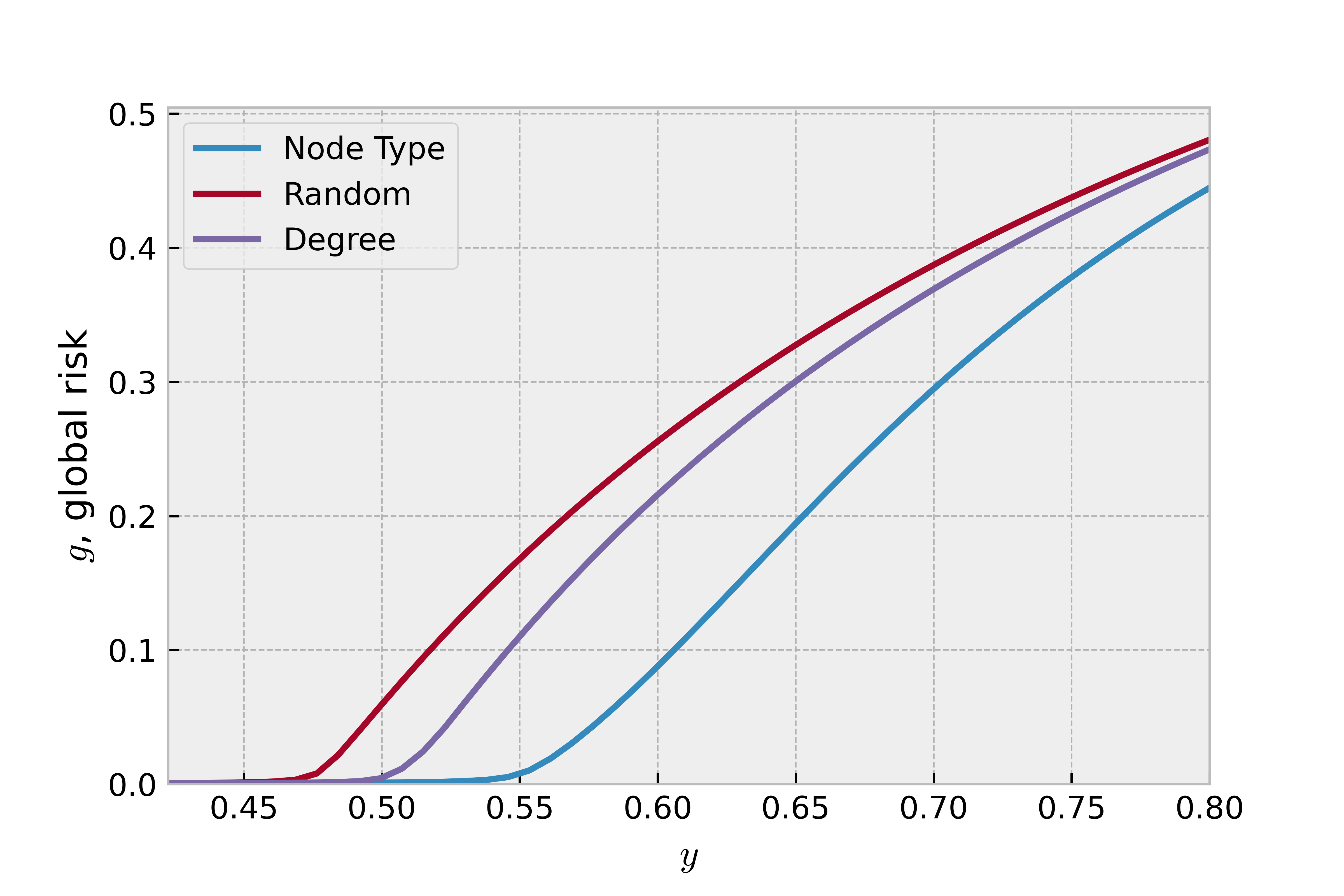}
		\caption{ Global risk plotted as a function of the fraction $y$ of links removed from an ER graph with mean degree $\kav=5$. Nodes have infectious time distribution $\gamma(t|\xi)= \xi \rme^{-\xi t}$ and are assigned to one of two sub-types of slow and fast recovery, $\xi \in \{\xi^{\ell},\xi^{h}\} = \{1,5\}$ such that $\xi^{\ell} < \xi^{h}$, with probability $\mathrm{P}(\xi^{h}) = 1- \mathrm{P}(\xi^{\ell}) = 0.65$. The infection rate is set to $\beta=2$. Results are shown for the cases where links are removed i) at random ii) with preference for links connecting nodes of high degree $y= \sum_{k,k'}\mathrm{W}(k,k')\alpha \left(1-\frac{k k'}{k_{\max}}\right)$, such that $\alpha$ set the value of $y$, and iii) with preference for nodes with slow recovery $y= \sum_{\xi,\xi'}\mathrm{W}(\xi;\xi')y(\xi,\xi')$ where we set $y(\xi^{h},\xi^{h})=1$ and $y(\xi^{\ell},\xi^{h}) = y(\xi^{\ell},\xi^{\ell})$ such that $y(\xi^{\ell},\xi^{\ell})$ is the free parameter that controls $y$.
		}
		\label{fig: lockdown strategies}
	\end{figure}

	We consider three different types of bond percolation: random percolation $y_{ij}=y ~\forall~ i,j$, degree-based percolation $y_{ij} = y(k_{i},k_{j})$, and sub-type  percolation $y_{ij} = y(\xi_{i},\xi_{j})$. Following steps in \ref{app:hetero}, one can find the closed expressions for the global risk
	\begin{eqnarray}
		\hspace{-0.5cm} g &= 1 - \sum_{k \xi}\mathrm{P}(k,\xi)\int \rmd t\gamma(t|\xi) \left(1 - \alpha(t) \tilde{g}_{k,\xi}\right)^{k}\\
		\hspace{-0.5cm}   \tilde{g}_{k,\xi} &=  \sum_{k'\xi'}y(k,\xi,k',\xi')\mathrm{W}(k',\xi'|k,\xi)\left[ 1 - \int \rmd t\gamma(t|\xi') \left(1 - \alpha(t) \tilde{g}_{k',\xi'}\right)^{k'-1}\right]
	\end{eqnarray}
	where $y(k,\xi,k',\xi')= y$ for random bond percolation,  $y(k,\xi,k',\xi')= y(k,k')$ for degree-dependent bond percolation and $y(k,\xi,k',\xi')= y(\xi,\xi')$ for sub-type dependent bond percolation. 
	In figure \ref{fig: lockdown strategies} we consider an ER network, with individuals randomly assigned one of two sub-types, high and low transmissibility, $\xi \in \{\xi^{\ell},\xi^{h}\}$ with $\xi^{\ell}<\xi^{h}$, such that degree and node sub-type are uncorrelated. We then compare the global risk following random, degree-based, and sub-type bond percolation for the same fraction of links removed. For degree-based percolation we choose  $y(k,k') =  \alpha\left(1 - \frac{ k k'}{k_{\max}^{2}}\right)$ to preferentially remove links attached to nodes of high degree. For sub-type percolation, we consider the scenario where nodes of low transmissibility (i.e. fast recovery $\xi=\xi^{h}$) do not reduce their social contact, i.e. $y(\xi^{h},\xi^{h})=1$, and we set $y(\xi^{\ell},\xi^{h})=
	y(\xi^{\ell},\xi^{\ell})$, such that $y(\xi^{\ell},\xi^{\ell})$ is the only
	free parameter which controls the overall fraction of links removed. 
	We see that link percolation based upon degree and node-type yields lower global risk relative to random link percolation. Furthermore, for this parameterisation we see that percolation based upon node-type is preferable to degree-based percolation. This is due to the fact that degree-based percolation deletes a high concentration of links between nodes of low transmissibility, as compared to percolation based upon sub-types.

	\section{Beyond the mean: distribution of risk in the SIR model}\label{sec: dist of risk}

	\begin{figure}[t]
		\includegraphics[width=0.49\textwidth]{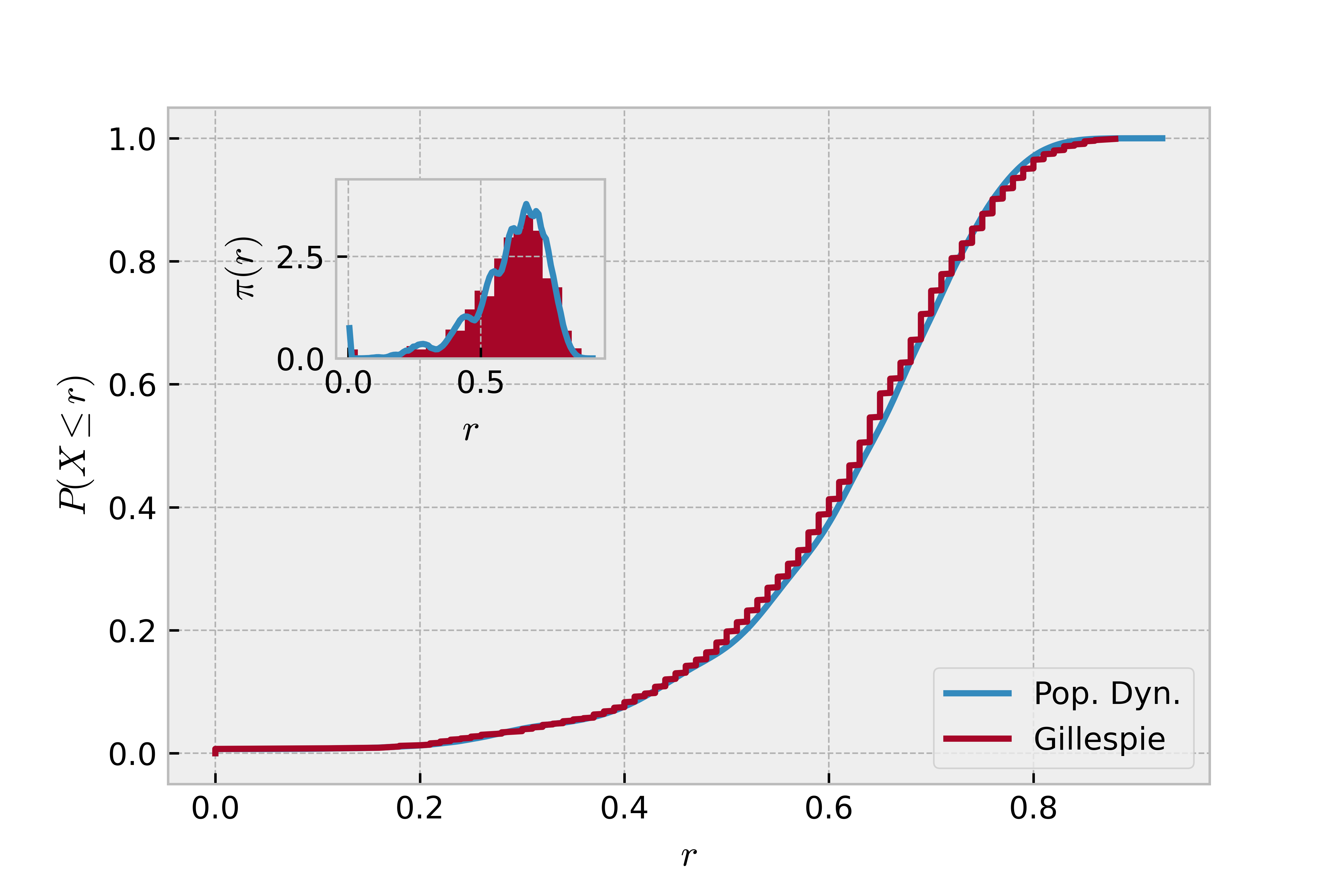}
		\includegraphics[width=0.49\textwidth]{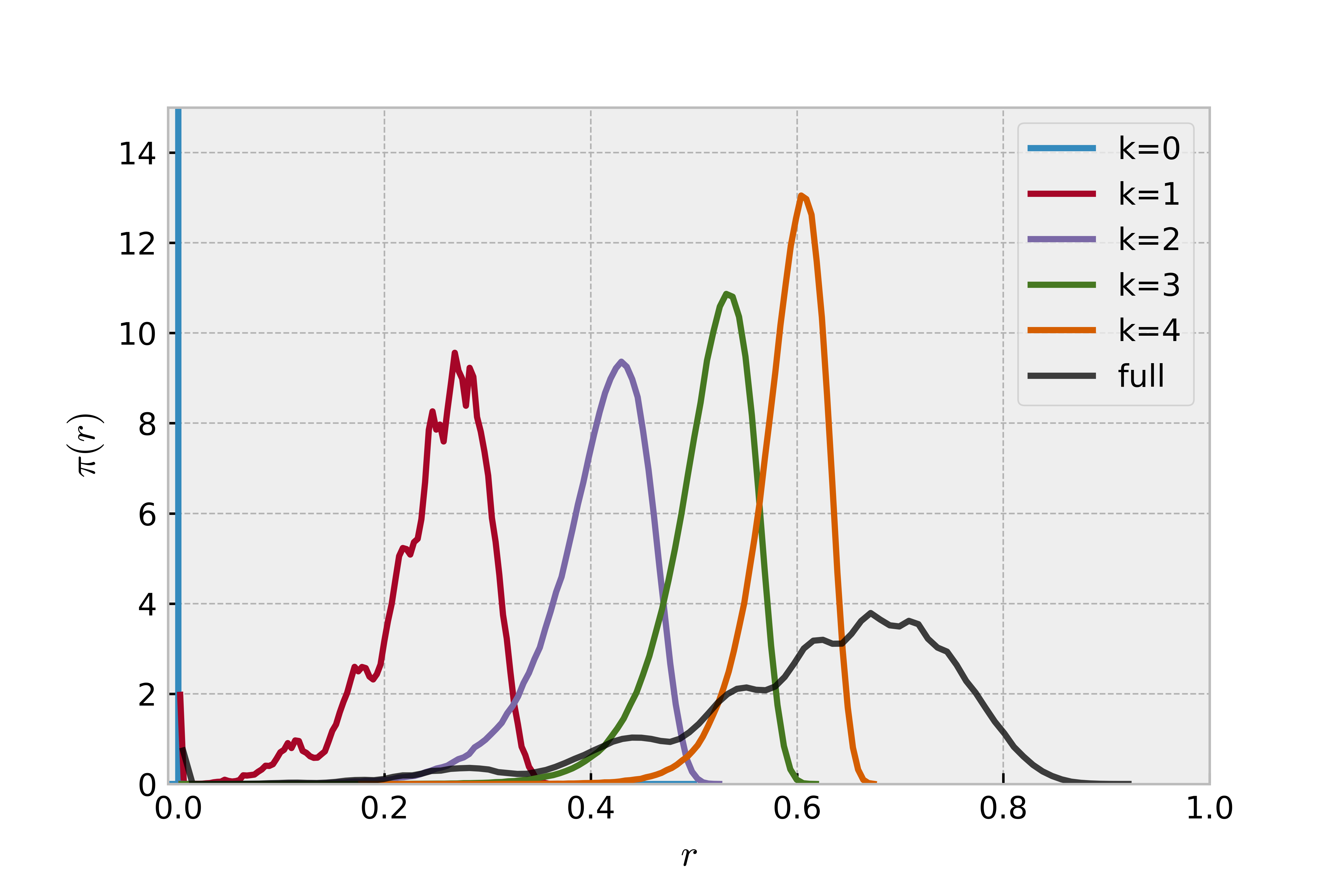}
		\caption{Results from population dynamics, with population size $S=2 \times 10^{5}$ and $2 \times 10^{6}$ samples to form the histogram, for an Erd\"os-R\'enyi network with mean degree $\kav=5$, infection rate $\beta=0.5$ and homogeneous transmission. Left: Cumulative
			distribution function of the risk, $\mathrm{P}(X \leq r)$, for inverse mean infectious time $\xi = 0.75$. Results from population dynamics and simulations on a network of size $N=2000$ are shown. Inset shows same information for the risk distribution $\pi(r)$, with simulations plotted as a histogram and population dynamics plotted as a solid line. Right: Distribution of the risk distribution $\pi(r)$ (black line) and degree conditional risk distributions $\pi_{k}(r)$, for $k\in\{0,1,2,3,4\}$ (peaks from left to right) for inverse mean infectious time $\xi = 0.6$.}. \label{fig: homo risk dist numerics}	
	\end{figure}

	\subsection{Distribution of risk in networks with degree correlations}
	
	In the previous sections we have used the cavity method to obtain a closed set of equations for the average risk of epidemics in a network where nodes have heterogeneous infectious time. 
	Previous work has shown that epidemic risk varies considerably from node to node, even with homogeneous transmission probabilities, due to variations in the local environment \cite{Rogers2015}. The presence of heterogeneities in 
	node infectious times is expected to add an extra source of fluctuations, therefore studying the distribution of the node risks is expected to provide important information that is not 
	captured by the mean risk.  
	
	Recently, the cavity method has been used to assess the heterogeneous behaviour of individual nodes in the context of percolation in sparse networks \cite{Rogers2015,Kuhn2017}. 
	Here we extend this approach to account for degree correlations and heterogeneities 
	in the node transmissability and calculate the functional form 
	of the distribution of node risks. 
	The latter
	is defined as
	\begin{eqnarray}
		\pi(r |\bA) = \frac{1}{N}\sum_{i=1}^N\delta\left(r- r_{i}(\bA)\right) \label{eq: def risk dist}
	\end{eqnarray}
	and a closed set of equation can be derived for it by 
	substituting the RHS of equation (\ref{eq:r}) into equation (\ref{eq: def risk dist}) and following a similar procedure as in \ref{app:homo} for the global risk, which leads, after averaging over the graph ensemble, to
	\begin{eqnarray}
		&\pi(r) = \sum_{k\geq0}\mathrm{P}(k) \left\{\prod_{j=1}^{k}\int \rmd\hr_{j}\hat{\pi}_{k}(\hr_{j})\right\}\delta\left(r- \phi(k,\boldsymbol{\hr})\right) \label{eq:pi} \\
		&\hat{\pi}_{k}(\hr) = \sum_{k'\geq1}\mathrm{W}(k'|k) \left\{\prod_{j=1}^{k'-1}\int \rmd\hr'_{j}\hat{\pi}_{k'}(\hr'_{j})\right\}\delta\left(\hr- \phi(k'-1,\boldsymbol{\hr}')\right) \label{eq:pi-hat}
	\end{eqnarray}
	with $\boldsymbol{\hr} = \{\hr_{1},...,\hr_{k} \}$, ${\bf \hr}'=(\hr'_1,\ldots,\hr'_{k'-1})$ and 
	\begin{eqnarray}
		\phi(k,\boldsymbol{\hat{r}}) = 1 - \int_{0}^{\infty}\rmd t\gamma(t) \prod_{j=1}^{k}\left(1 - \alpha (t)\hat{r}_{j}\right).
	\end{eqnarray}
	The set of equations
	(\ref{eq:pi-hat}) for the degree-dependent `cavity' risk distributions $\hat{\pi}_{k}(\hat{r})$ 
	must be solved first to then solve equation (\ref{eq:pi}) for the distribution of risk $\pi(r)$. This can be accomplished numerically by using a population dynamics algorithm \cite{Mezard2001}.
	This consists of approximating each $\hat{\pi}_{k}(\hr)$ by the empirical cavity risk frequencies computed from a large number (i.e. a
	population) of cavity risks, which are updated at each iteration of the algorithm following a stochastic map. We start by initializing 
	$k_{\max}$ populations of cavity risks, each of size $S$, by drawing $k_{\max} \times S$ random variables 
	$\hr_i^{(k)}$, $i=1,\ldots, S$ in the $[0,1]$ interval.
	Their empirical distribution represents the
	zero-step approximation $\hat{\pi}_{k}^{0}(\hr)$ of $\hat{\pi}_{k}$. We then evolve the risks such that at each step $n$ the empirical distribution 
	$\hat{\pi}_{k}^{n}(\hr)$ comes closer and closer to the invariant distribution $\hat{\pi}_{k}(\hr)$. Each iteration of the algorithm consists of the following steps, which are repeated until convergence 
	\begin{enumerate}
		\item Set $k=1$. 
		\item Draw degree $k'$ with probability  $\mathrm{W}(k'|k)$ \label{step: draw k'}.
		\item Randomly select $k'-1$ risks from the $k^{\prime}$-th population of risks. \label{step: select individ}
		\item Compute the function $\phi(k'-1,\boldsymbol{\hat{r}})$ using the $k'-1$ values of risk.
		\item Select a random risk from the $k$-th population and set its value to $\phi(k'-1,\boldsymbol{\hat{r}})$. 
		\label{step: set val2}
		\item Set $k=k+1$ and go back to step (\ref{step: draw k'}). 
		\item Repeat all steps until all populations have converged.
	\end{enumerate}
	Once the populations have converged, i.e. $\hat{\pi}_{k}(\hat{r})$ does not change upon further iteration, the distribution of risk $\pi(r)$ can be computed in a similar manner by taking samples from the population and computing the histogram of $\phi(k,\boldsymbol{\hat{r}})$. 
	
	In Figure \ref{fig: homo risk dist numerics} (left panel) we show results from the population dynamics algorithm for an ER graph with average connectivity $\kav=5$. Results are in very good agreement with simulations, carried out on networks of size $N=2000$.  As previously shown in \cite{Kuhn2017}, the distribution of risk can be conditioned upon the degree of a node, leading to the degree-conditional risk distributions
	\begin{eqnarray}
		&\pi_{k}(r) =  \left\{\prod_{j=1}^{k}\int \rmd\hat{r}_{j}\hat{\pi}_{k}(\hat{r}_{j})\right\}\delta\left(r- \phi(k,\boldsymbol{{\hat{r}}})\right). \label{eq: Homo risk dist degree decon}
	\end{eqnarray}
	These are plotted in figure \ref{fig: homo risk dist numerics} (right panel) and show that the peaks and troughs apparent in $\pi(r)$ are caused by nodes of different degree, with the peak at $r=0$ arising from disconnected nodes. From this figure we see that the mean risk increases with degree, showing the benefit of degree-based vaccination strategies. However, the skew in the degree-conditional risk distributions
	means that degrees which have a high \emph{mean} risk, also contain nodes of relatively low risk, and vaccinating these nodes would reduce the overall risk and epidemic size by little, highlighting the limitations of a degree-based strategy. It can be checked analytically and verified numerically that the risk distribution for random regular graphs are delta-peaked, with the peak corresponding to the solution to equation (\ref{eq:ghat-W}) for the global risk, which suggests that the random nature of the degree of the neighbours of a node in graphs with heterogeneous structure are the source of variability in the risk of nodes with the same degree.

\begin{figure}[t]
		\includegraphics[width=0.49\textwidth]{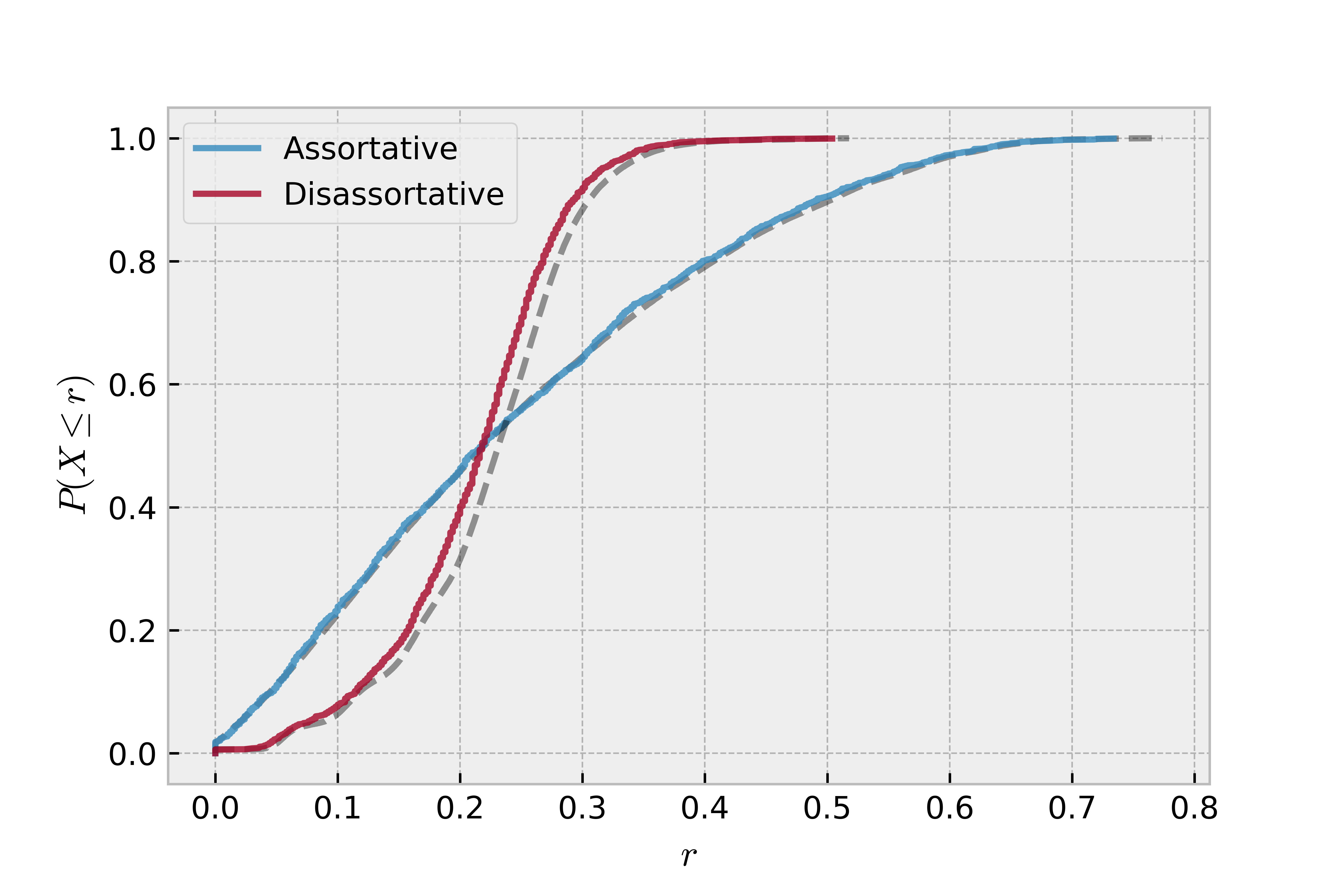}
		\includegraphics[width=0.49\textwidth]{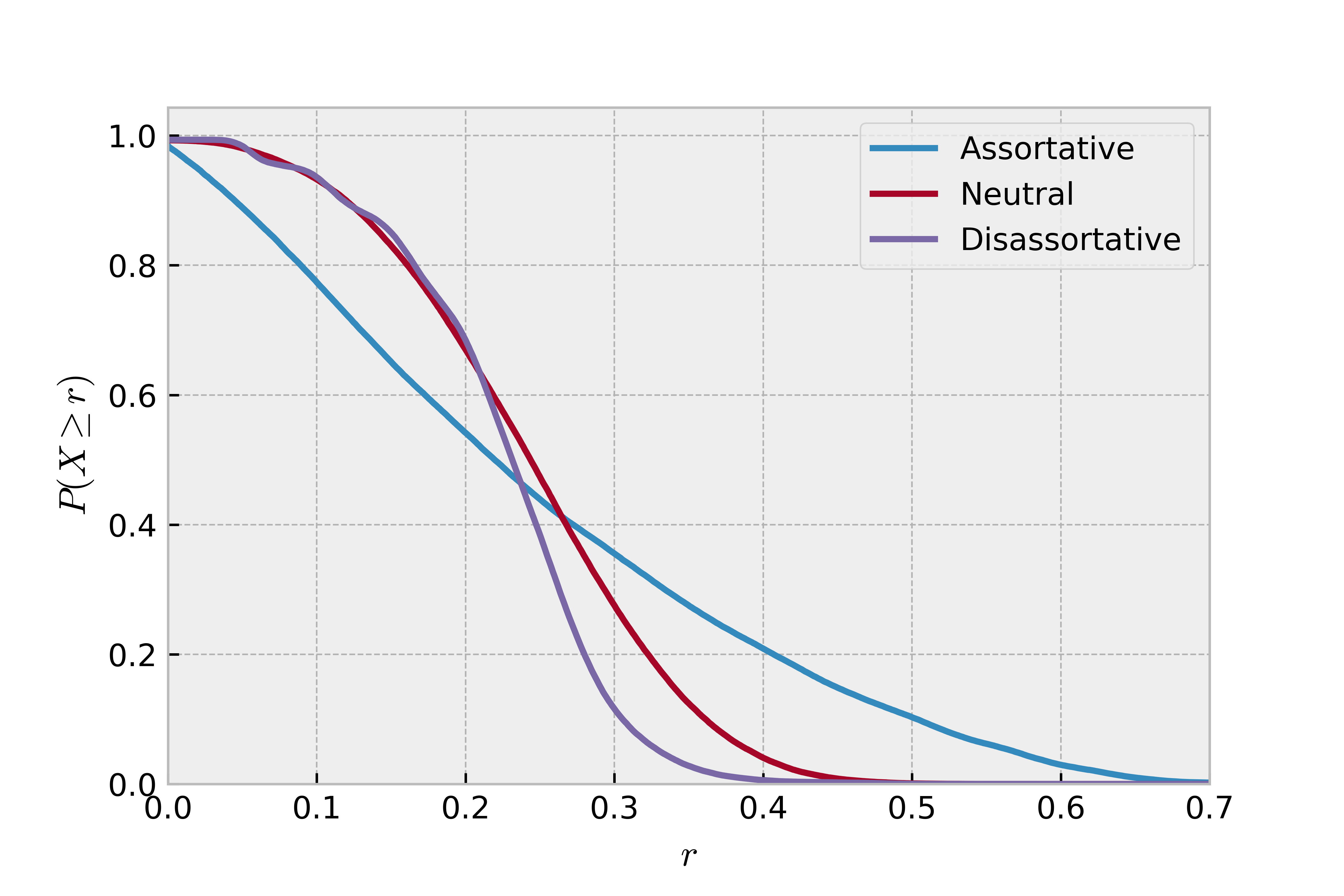}
		\caption{Results from population dynamics, with population size $S=1 \times 10^{4}$ and $1 \times 10^{5}$ samples to form the histogram, for a network with Poissonian degree distribution, mean degree $\kav=5$, infection rate $\beta=0.25$ and homogeneous recovery rate $\xi=0.75$. Left:
			cumulative distribution function of the risk, $\mathrm{P}(X\leq r)$, computed from the cavity method (dashed line) compared with simulations (solid line) on a network of size $N=2000$, for assortative and disassortative degree correlations. Right: same information from the cavity method shown for the survival function, $\mathrm{P}(X \geq r)$, with the curve for a neutral graph added for comparison} \label{fig: risk dist degree correlations}	
	\end{figure}

	Figure \ref{fig: risk dist degree correlations} (left panel) shows the cumulative distribution of risk for 
	networks with Poissonian degree distribution and assortative and disassortative degree correlations, respectively. 
	Results from the cavity method are found in good agreement with simulations on networks with size $N=2000$. Degree correlations are seen to have a significant impact on the risk distribution.  
	This is further shown on the right panel, where the survival function $\mathrm{P}(X\geq r)$ of assortative graphs is seen to have larger tails than disassortative or neutral graphs, due to a higher proportion of high risk nodes.
	
	\subsection{Distribution of risk in networks with heterogeneous transmission}
\begin{figure}[t]
		\includegraphics[width=0.49\textwidth]{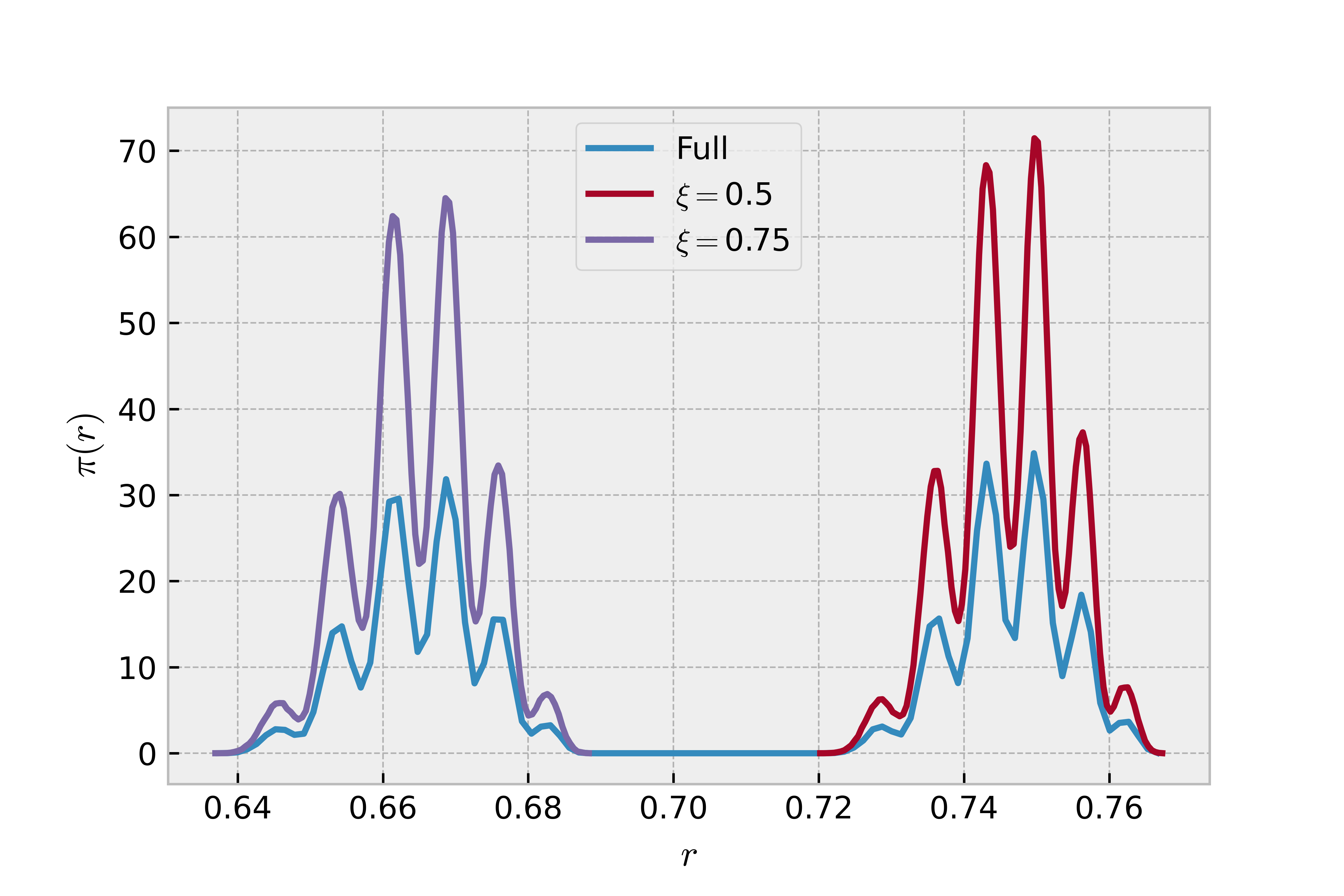}
		\includegraphics[width=0.49\textwidth]{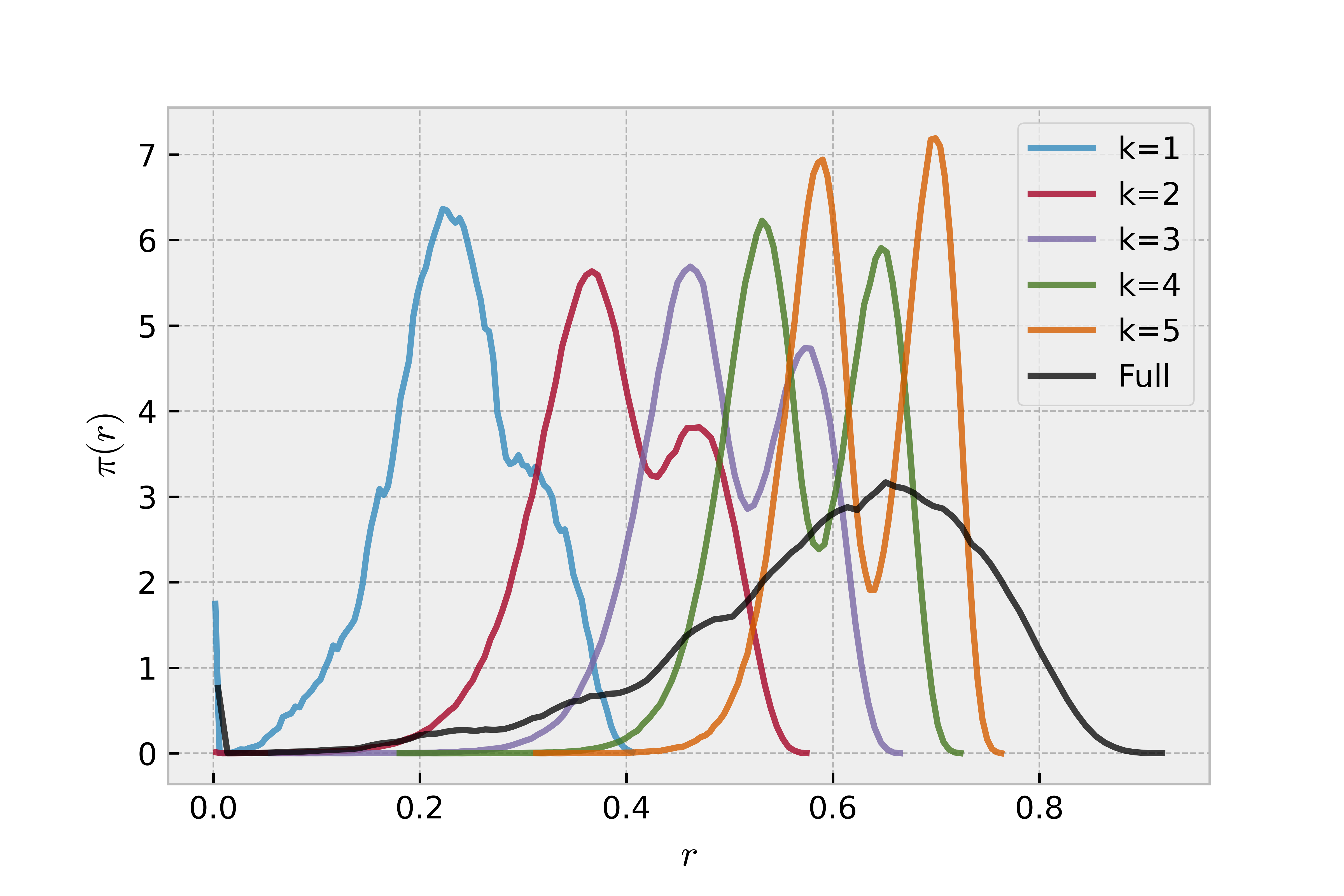}
		\caption{Left: Risk distribution $\pi(r)$ of regular graph with connectivity $\kav=5$, rate of infection $\beta =0.5$, and two recovery rates $\xi \in\{ 0.5,0.75\}$ with probability $\mathrm{P}(\xi)=0.5$ for all $\xi$. The curves with larger peaks correspond to the risk distribution conditioned upon $\xi$. Right: risk distribution $\pi(r)$ for a network from the ensemble (\ref{eq: ModER ensemble}) with mean degree $\kav=5$, rate of infection $\beta =0.5$, and two recovery rates $\xi \in\{ 0.6,1.0\}$ with probability $\mathrm{P}(\xi=1.0)=0.5$. The transmissability kernel is defined in (\ref{eq: two cluster kernel}) with $\bar{k}(\xi=0.6)=\bar{k}(\xi=1.0)=\kav$. Larger peaks correspond to the risk distribution conditioned upon $k$. In both plots the population size is $S=10^{5}$ and $10^{6}$ samples are taken from the converged cavity distribution to form $\pi(r)$.
		}
		\label{fig: hetero risk dist}
	\end{figure}

	The equations for the distribution of risk can be immediately generalised to the case of heterogeneous transmission, giving 
	\begin{align}
		\pi(r) &= \sum_{k\geq0,\xi}\mathrm{P}(k,\xi) \left\{\prod_{j=1}^{k}\int \rmd\hat{r}_{j}\hat{\pi}_{k,\xi}(\hat{r}_{j})\right\}\delta\left(r- \phi(k,\xi,\boldsymbol{{\hat{r}}})\right) \label{eq:pi-hetero} \\
		\hat{\pi}_{k,\xi}(\hat{r}) &= \sum_{k'\geq1,\xi'}\mathrm{W}(k',\xi'|k,\xi) \left\{\prod_{j=1}^{k'-1} \int \rmd\hat{r}'_{j}\hat{\pi}_{k',\xi'}(\hat{r}'_{j})\right\}\delta\left(\hat{r}- \phi(k'-1,\xi',\boldsymbol{{\hat{r}'}})\right)    \label{eq:pi-hat-hetero}\\
		\phi(k,\xi,\boldsymbol{\hat{r}}) &= 1 - \int_{0}^{\infty}\rmd t\gamma(t|\xi) \prod_{j=1}^{k}\left(1 - \alpha (t)\hat{r}_{j}\right).
	\end{align}
	These equations can be solved using a generalisation of the population dynamics algorithm above with $P \times k_{\max}$ populations. The risk distribution $\pi(r)$ can be de-convoluted either in terms of degree, or node label $\xi$ or both. To illustrate this, we first consider a random regular graph with degree $\kav=5$, where each node is independently assigned a random label $\xi \in \{\xi^\ell,\xi^h\}$. Figure \ref{fig: hetero risk dist} (left panel) shows that the risk distribution neatly splits into the distribution of risk for each cluster. 
	The risk distribution for nodes of a given label $\xi$, is seen to have several peaks, corresponding to the different neighbourhoods a node may have: in a regular graph of degree $c$, with two clusters, there are $c+1$ different labelling configurations for the neighbourhood of a node. The height of the peaks in this distribution are therefore related to the multiplicity of these possible labelling configurations. It is important to note, that these peaks are not delta-peaked, but are broader, and overlap, due to the risk being dependent not only upon the neighbourhood of a node, but on the neighbours of neighbours as explored in \cite{tishby2018revealing}. Figure \ref{fig: hetero risk dist} (right panel) shows the risk distribution and the degree conditional risk distributions for the ensemble defined in (\ref{eq: ModER ensemble}), with average connectivity $\kav=5$. When compared to figure \ref{fig: homo risk dist numerics} we see a greater amount of heterogeneity for nodes of a given degree. This highlights the importance of understanding differences in transmissability in individuals and its interplay with the network structure: the efficacy of sub-type based or degree based vaccination depends on the level of links between each cluster and their relative transmissability.

	\subsection{Risk distribution in the limit of large connectivity}
\begin{figure}[t]
		\includegraphics[width=0.49\textwidth]{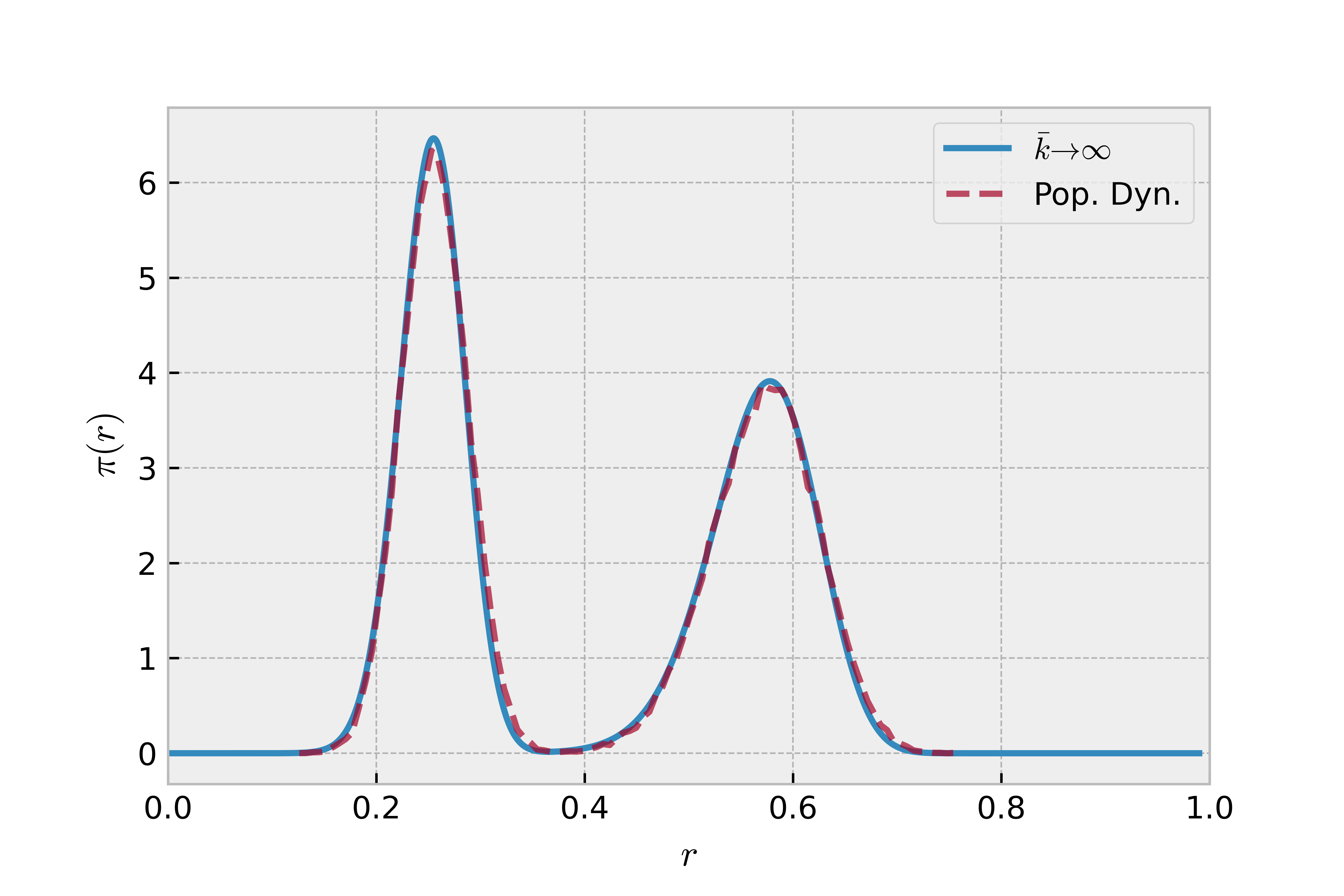}
		\includegraphics[width=0.49\textwidth]{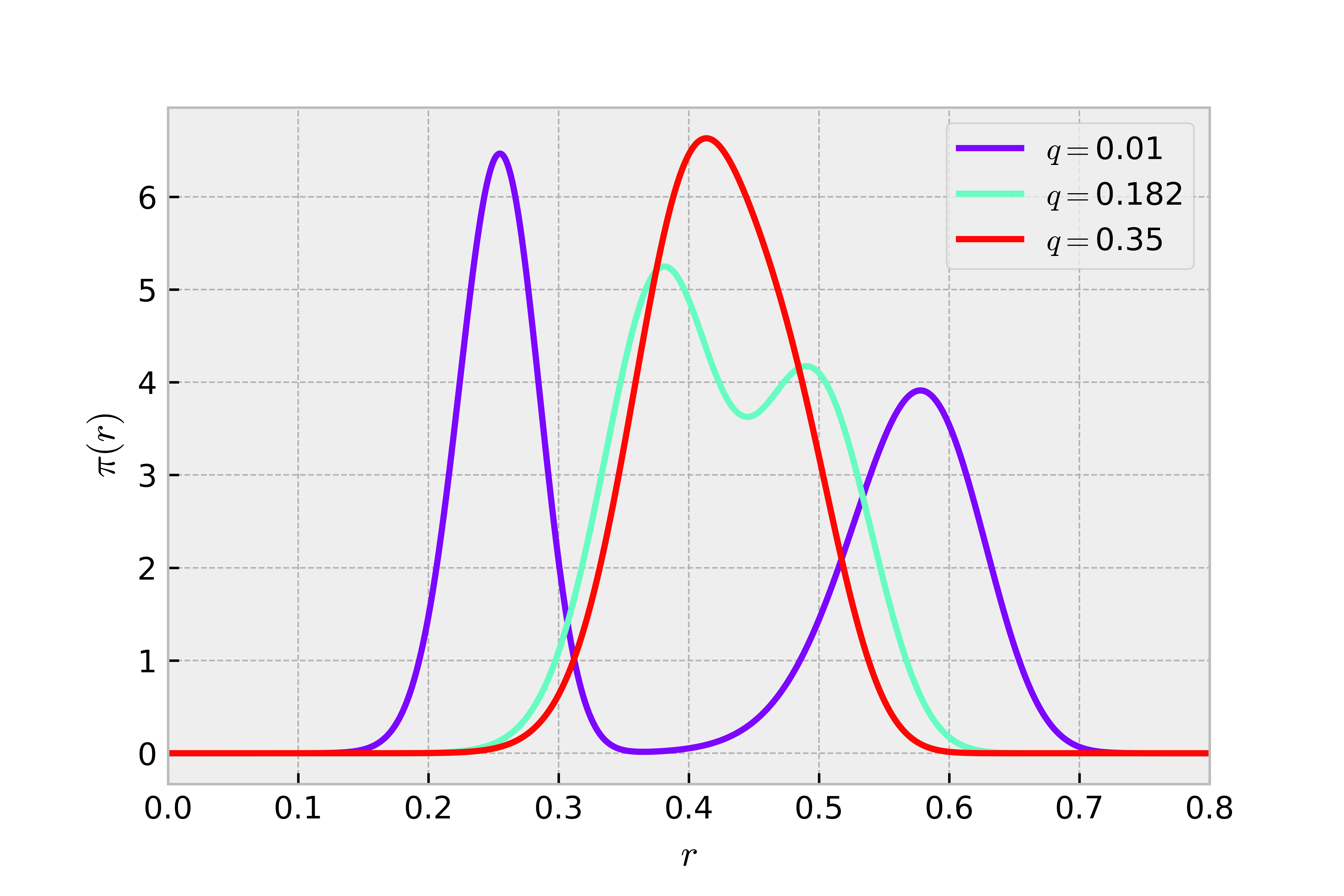}
		\caption{Conditional risk distribution $\pi_{\xi}(r)$, for a network in the ensemble (\ref{eq: ModER ensemble}) with $\kav=50$ and two clusters, with recovery time distribution $\gamma(t|\xi) = \delta(t + \frac{1}{\beta}\ln(1 - T(\xi)))$, parameterised by $T(\xi) \in \{0.0225,0.03\}$ with $\mathrm{P}(\xi)=0.5$ and inter-connectivity $q$.
			Left: The dashed lines show the solutions to equations (\ref{eq:pi-hetero}) and (\ref{eq:pi-hat-hetero}) via population dynamics, with population size $S=10^{5}$ and $10^{6}$ samples forming the final distribution. The solid curve shows the large mean limit approximation (\ref{eq: hetero large mean }). The inter-connectivity is set to $q=0.01$. 
			Right: Large mean limit approximation for different values of $q$, as shown in the legend.}	 \label{fig: hetero risk dist large mean}	
	\end{figure}

	An exact expression for the risk distribution may be derived in the limit of large connectivity, for graphs with homogeneous transmissibility, using methods demonstrated in \cite{Kuhn2017}. To extend these methods to graphs with heterogeneous transmissibility we use the {\em ansatz} that as $\kav \to \infty$ the cavity field distributions will become delta-peaked i.e. $\hat{\pi}_{k,\xi}(\hat{r}) = \delta(\hat{r} - \hat{r}^{*}_{k,\xi})$. 
	Inserting this into equation (\ref{eq:pi-hat-hetero}), multiplying by $\hat{r}$ and integrating both sides over $\hat{r}$, we find
	\begin{eqnarray}
		\hspace{-5mm}
		\hat{r}^{*}_{k,\xi} = \sum_{k',\xi'}\mathrm{W}(k',\xi'|k,\xi)\left[1 - \int_{0}^{\infty}\rmd t\gamma(t|\xi')\left(1 -\alpha(t)\hat{r}^{*}_{k',\xi'}\right)^{k'-1}\right]. \label{eq: rstar large mean lim}
	\end{eqnarray}
	Similarly, we insert $\hat{\pi}_{k,\xi}(\hat{r}) = \delta(\hat{r} - \hat{r}^{*}_{k,\xi})$ into equation (\ref{eq:pi-hetero}) which yields, 
	\begin{align}
		\pi(r) &= \sum_{k,\xi} \mathrm{P}(k,\xi)\delta\left(r - \left[1 - \int_{0}^{\infty} \rmd t\gamma(t|\xi) \left(1 - \alpha(t)\hat{r}^{*}_{k,\xi}\right)^{k} \right]\right).
	\end{align}
	Given that $\pi(r)=\sum_{k,\xi}\pi(r|k,\xi)\mathrm{P}(k,\xi)$ we find the conditional risk distribution to be delta-peaked under this ansatz,
	\begin{eqnarray}
		\pi(r|k,\xi) &= \delta\left(r - f_{\xi}(k)\right)
		\label{eq:delta_r}
	\end{eqnarray}
	where
	\begin{eqnarray}
		f_{\xi}(k) &= 1 - \int_{0}^{\infty}\rmd t\gamma(t|\xi) \left(1 - \alpha(t) \hat{r}^{*}_{k,\xi}\right)^{k}.
	\end{eqnarray}
	We can simplify the above by choosing $\gamma(t|\xi) = \delta\left(t + \frac{1}{\beta}\ln\left(1-T(\xi)\right)\right)$ such that, 
	\begin{eqnarray}
		f_{\xi}(k) &= 1 - \left(1 - T(\xi) \hat{r}^{*}_{k,\xi}\right)^{k}.
	\end{eqnarray}
	In order to retrieve a non-trivial distribution of risk in the limit $\bar{k}\to\infty$, 
	we assume that $T(\xi) \ll 1 ~ \forall ~ \xi$ (at finite transmissibility, every node will be near-certain of causing an epidemic in the large connectivity limit).  This allows us to write $f_{\xi}(k) = 1 - \rme^{- kT(\xi)\hat{r}^{*}_{k,\xi}}$. To proceed we assume that degree correlations are of the form $W(k',\xi'|k,\xi) = W(\xi'|\xi)W(k'|\xi')$, which by equation (\ref{eq: rstar large mean lim}) yields $r^{*}_{k,\xi} = r^{*}_{\xi} ~ \forall ~ k$. 
	To find the risk distribution of nodes of a given label $\pi_{\xi}(r)=\sum_{k}\pi(r|k,\xi)\mathrm{P}(k|\xi)$, we then use (\ref{eq:delta_r}) and properties of the Dirac delta-function to write, 
	\begin{eqnarray}
		\hspace*{-2cm}    \pi_{\xi}(r)&= \sum_{k}\mathrm{P}(k|\xi)\frac{\delta \left(k - f^{-1}_{\xi}(r)\right)}{|f'_{\xi}(f^{-1}_{\xi}(r))|}=\frac{1}{T(\xi)\hat{r}_{\xi}^*(1-r)}\sum_{k}\mathrm{P}(k|\xi)\delta\left(
		k+\frac{\ln(1-r)}{T(\xi)\hat{r}_{\xi}^*}
		\right). \label{eq: large mean limit conditional general}
	\end{eqnarray}
	To proceed further we must specify the degree distribution, and so we consider the distribution of risk in graphs drawn from the graph ensemble defined by (\ref{eq: ModER ensemble}) and assume that as $\kav \to \infty$ the conditional mean degree $\bar{k}(\xi) \to \infty ~ \forall ~ \xi$. Noting that the degrees in this ensemble are distributed according to $\mathrm{P}(k|\xi) = \rme^{-\bar{k}(\xi)}\Bar{k}^k(\xi)/k!$, which in the limit of $\Bar{k}(\xi) \to \infty$ is well approximated by $\mathrm{P}(k|\xi) \sim \mathcal{N}(\Bar{k}(\xi),\Bar{k}(\xi))$, we find, 
	\begin{eqnarray}
		\pi_{\xi}(r) = \frac{1}{T(\xi) r^{*}_{\xi}} \sqrt{\frac{1}{2 \pi \bar{k}(\xi)}} \exp\! \left [\! - \frac{\bar{k}(\xi)}{2}\left(\!1 + \frac{\ln(1 - r)}{T(\xi)r^{*}_{\xi}\bar{k}(\xi)}\!\right)^{2}\!\! - \ln(1-r)\! \right]. \label{eq: hetero large mean }
	\end{eqnarray}
	The unconditional risk distribution is then given by $ \pi(r) = \sum_{\xi}\mathrm{P}(\xi) \pi_{\xi}(r)$. We can compare this expression with the results of population dynamics for a system with two clusters with transmissability $T(\xi) = \{0.0225,0.03\}$ of equal size $\mathrm{P}(\xi) = 0.5$. We see in figure \ref{fig: hetero risk dist large mean} that for the case where $\kav=50$ there is excellent agreement with the expression (\ref{eq: hetero large mean }). In figure \ref{fig: hetero risk dist large mean} we also see the impact of changing the inter-connectivity in this ensemble: for low $q$ we see two distinct peaks, corresponding to the risk of nodes of different transmissability, but as $q$ increases the two peaks begin to overlap, until the distribution becomes unimodal for high $q$.  
	
	\subsection{Distributional equations of risk with node and link percolation}
	
	Finally, we use the cavity method to investigate the effect of node and link percolation on the distribution of risk in the SIR model. Node deletion models perfect vaccination, as before, and link percolation models the loss of a social contact between two individuals, due to, for example, social-distancing measures. To do so we introduce a binary variable $\tau_{ij}$ which indicates whether a link in a contact network has been deleted ($\tau_{ij}=0$) or not ($\tau_{ij}=1$). The local risk subject to node and link percolation is then written, 
	\begin{eqnarray}
		r_{i}(\bA) &=\sigma_{i}\left[1 - \int_{0}^{\infty}\rmd t\,\gamma(t)\prod_{j \in \partial^\bA_{i}}\left(1 - \tau_{ij}\alpha(t) r_{j}^{(i)}(\bA )\right)\right] \label{eq:r-link} \\ 
		r_{j}^{(i)}(\bA) &= \sigma_{j}\left[ 1 - \int_{0}^{\infty}\rmd t\,\gamma(t)\prod_{\ell \in \partial^\bA_{j}\setminus i}\left(1 - \tau_{jl}\alpha(t) r_{\ell}^{(j)}(\bA)\right)\right]. \label{eq:rhat-link}
	\end{eqnarray}
	We consider the case where node and links are randomly and independently deleted with probability dependent on their degree, so that 
	\begin{eqnarray}
		\mathrm{P}(\bsigma|\bk)  &= \prod_{i}\left[(1 - x(k_{i}))\delta_{\sigma_{i},0} + x(k_{i})\delta_{\sigma_{i},1}\right]\label{eq: degree node deletion} \\
		\mathrm{Q}(\btau|\bk)  &= \prod_{i<j}\left[(1 - y(k_{i},k_{j}))\delta_{\tau_{ij},0} + y(k_{i},k_{j})\delta_{\tau_{ij},1}\right]. \label{eq: degree link deletion}
	\end{eqnarray}
	By proceeding as in \ref{app:homo}, and averaging over the graph ensemble, we find that the distribution of risk is given by, 
	\begin{eqnarray}
		\pi(r) &=& \sum_{k}\mathrm{P}(k)\sum_{\sigma}P(\sigma|k)\nonumber\\
		&&\times\left\{\prod_{j=1}^{k} \sum_{k_{j}'}\mathrm{W}(k_{j}'|k) \sum_{\hat{\tau}_{j}}\mathrm{Q}(\hat{\tau}_{j}|k,k_{j}')\int \rmd\hr_{j}\hat{\pi}_{k_{j}'}(\hr_{j})\right\}\nonumber\\
		&&\times\delta\left(r- \phi(k,\sigma,\boldsymbol{\hat{\tau}},\boldsymbol{\hr})\right) \nonumber \\
		\hat{\pi}_{k}(\hr) &=&\sum_{\sigma}\mathrm{P}(\sigma|k)\left\{\prod_{\ell=1}^{k-1} \sum_{k_{\ell}}\mathrm{W}(k_{\ell}'|k) \sum_{\hat{\tau}'_{\ell}}\mathrm{Q}(\hat{\tau}'_{\ell}|k,k_{\ell}')\int \rmd\hr_{\ell}'\hat{\pi}_{k_{\ell}'}(\hr_{\ell}')\right\}\nonumber\\
		&&\times\delta\left(\hr- \phi(k-1,\sigma,\boldsymbol{ \hat{\tau}'},\boldsymbol{\hr'})\right) \label{eq:pi-hat-percolation}
	\end{eqnarray}
	with $\boldsymbol{\hat{\tau}} = \{\hat{\tau}_{1},...,\hat{\tau}_{k} \}$, ${\boldsymbol{ \hat{\tau}}'}=(\hat{\tau}'_1,\ldots,\hat{\tau}'_{k'-1})$ and
	\begin{eqnarray}
		\phi(k,\sigma,\boldsymbol{\hat{\tau}},\boldsymbol{\hat{r}}) = \sigma\left[1 - \int_{0}^{\infty}\rmd t\gamma(t) \prod_{j=1}^{k}\left(1 - \alpha (t)\hat{\tau}_{j}\hat{r}_{j}\right)\right].
	\end{eqnarray}
	Equation (\ref{eq:pi-hat-percolation}) can be solved by a generalisation of the populations dynamics algorithm above. To evaluate the cavity distribution $\hat{\pi}(\hr)$: 
	\begin{enumerate}
		\item Set $k=1$. 
		\item Draw $\sigma$ with probability $\mathrm{P}(\sigma|k)$
		\item Draw $k'$ degree from the distribution $\mathrm{W}(k'|k)$ \label{step: percolation draw k'}.
		\item Select a risk from the $k^{\prime}$-th population of risks. \label{step: percolation select individ}
		\item Select $\hat{\tau}'$ with probability $\mathrm{Q}(\hat{\tau}'|k,k')$. \label{step: percolation select tau } 
		\item Repeat steps (\ref{step: percolation draw k'}-\ref{step: percolation select tau }) $k-1$ times, and store each $\tau$, and individual to form the vectors $\boldsymbol{\hat{\tau}'}$ and $\boldsymbol{\hat{r}'}$.
		\item Select a random risk from the $k$-th population and set its value to $\phi(k-1,\sigma,\boldsymbol{\hat{\tau}'},\boldsymbol{\hat{r}'})$.\label{step: set val}
		\item Set $k=k+1$ and go back to step (\ref{step: draw k'}). 
		\item Repeat all steps until all populations have converged.
	\end{enumerate}
	Once the populations estimating $\hat{\pi}_{k}(\hr)$ have converged, one can compute the risk distribution in a similar manner. We note that, by equations (\ref{eq:pi-hat-percolation}), the risk distribution $\pi(r)$ can be deconvoluted in terms of $k$ and $\sigma$. By conditioning upon $\sigma=1$ it is possible to retrieve the risk distribution of unvaccinated nodes, therefore providing a method to see how vaccination strategies affect risk in the network of unvaccinated nodes. 
	
\begin{figure}[t]
		\centering
		\includegraphics[width=0.55\textwidth]{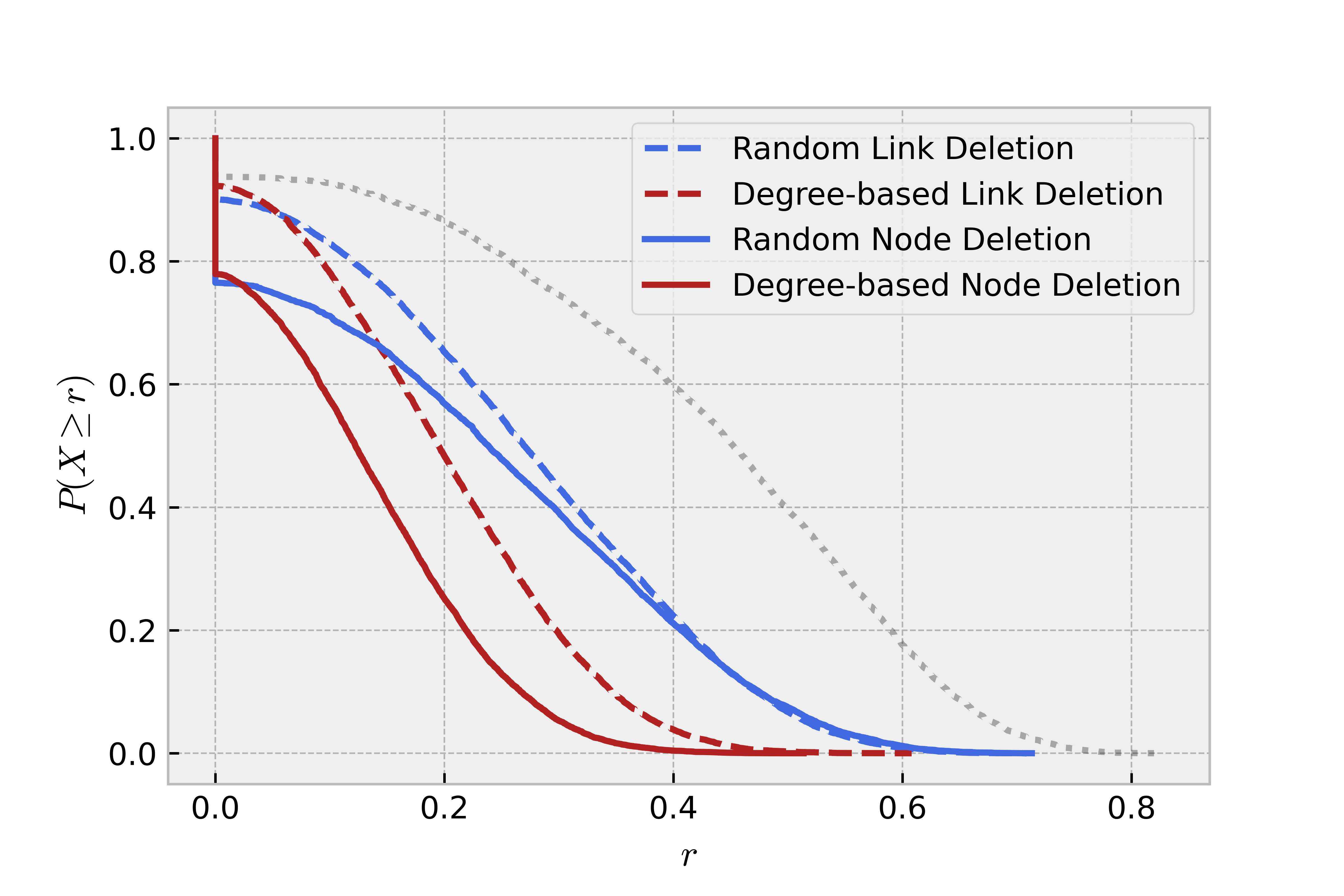}
		\caption{Survival function for the risk of nodes in a Erd\"os-R\'enyi graph with mean degree $\kav=3$. Infection rate is $\beta=0.7$ and nodes have exponential infectious time distribution $\gamma(t)=\xi \rme^{-\xi t}$ with mean infectious time $\xi=0.75$. The dotted line indicates case where no node or link deletion has occurred. In all other cases $15\%$ nodes/links are deleted, subject to either random (blue curves) or degree-based (red curves) deletion where high degree nodes/links are preferentially selected for deletion. For degree based node deletion we have nodes deleted according to (\ref{eq: degree node deletion}) with $x(k) = 1 - \alpha \frac{k}{k_{\max}}$, and for degree based link deletion we delete nodes according to (\ref{eq: degree link deletion}) with $y(k,k') =1 -  \alpha\frac{k k'}{k_{\max}^{2}}$}	 \label{fig: lockdown dist}	
	\end{figure}

	In figure \ref{fig: lockdown dist} we show the effects of node and link deletion, representing perfect vaccination and social distancing respectively. We consider two cases: random and degree-based strategies. For random vaccination we have $x(k)=x~ \forall ~k$ and $y(k,k')=y ~\forall~ k,k'$. For degree based strategies we set $x(k) =1- \alpha \frac{k}{k_{\max}}$ and $y(k,k') = 1-  \alpha \frac{k k'}{k_{\max}^{2}}$ such that nodes of high degree and links connecting nodes of high degree are more likely to be deleted. We show the case where $15\%$ of nodes/links are deleted in an ER graph. We see that for the same fraction of nodes/links deleted, the overall risk of the population is lower under random node deletion than random link deletion. Furthermore, risk under degree-based node deletion is lower than degree-based link deletion. This is consistent with the idea that social distancing measures, which assume an imperfect break in the chain of social contacts, are a less effective method of epidemic prevention than vaccination. It is however, interesting to compare the effect of degree-based link deletion, to random node deletion: in this case link deletion has the greater reduction of risk in the population. This highlights the importance of strategy in vaccination, although deleting a node blocks all paths of infection through that node, a targeted strategy of link deletion can lead to greater reductions in risk than random node deletion.

	\section{Discussion and conclusion}

	Predicting the effect of vaccination in epidemic models on networks continues to be a source of interesting inquiry. In our work we have relaxed the common assumption that vaccination blocks all transmission and instead assumed that vaccinated individuals have reduced transmissibility. From a modelling perspective this is useful when assessing the risk of epidemics and there is a lack of data for the impact of a given vaccine on transmission. To explore this, we have extended the cavity method to account for heterogeneity in the transmissibility of groups of individuals. These groups could represent differences in age, pre-existing medical conditions, or any other known correlates with transmissibility. Here however we have focused on a population of vaccinated and unvaccinated individuals, distinguished from each other by their transmissibility. Our results reveal that herd immunity is still attained from vaccination with partial transmission, if vaccination reduces transmissibility to a sufficiently low level that depends on both the infectious rate of the disease and contact network topology. We have shown that vaccination with partial transmission requires a greater proportion of the population to be vaccinated to achieve herd immunity, as intuitively expected. This highlights the importance of estimating the transmissibility of an infectious disease, and the impact that vaccination has upon individual transmissibility. However, it also highlights the robustness of vaccination as a strategy for the mitigation of epidemics, as even a partial reduction of transmission can achieve the desired herd immunity effect. 
	
	The benefit of the cavity approach to the SIR model on networks is that it allows one to evaluate the epidemic risk without simulation. Our equations for the global risk provide a quick method to calculate epidemic risk in populations with groups of different transmissibility. By solving the distributional equations of epidemic risk we have revealed the stark impact of heterogeneity in transmissibility. For graphs with homogeneous degree structure, the distribution of risk is delta-peaked when transmissibility of each individual is the same, but non-trivial when nodes take one of two values of transmissibility. This reveals that it is a strong assumption to make that the transmissibility of individuals is homogeneous, and that to ignore such heterogeneity can potentially dampen epidemic mitigation strategies that focus on network topology alone. Indeed our equations show that by introducing groups of different transmissibility, the epidemic risk will depend upon the size and transmissibility of each group, as well as the assortativity of links between nodes in different groups. 
	
	While vaccination is an important part of epidemic mitigation, we have also shown that the cavity method can model social distancing via bond percolation. Our framework allows us to compare random bond percolation with percolation that targets either links connecting nodes of high degree, or nodes of a particular sub-type. This may be useful when studying variations in the social distancing of different demographics and their impact on epidemic risk. One example could be studying the effect of relaxed social distancing amongst the vaccinated population who are perhaps less risk averse. 
	
	When studying populations with groups of different transmissibility we have provided a method to reduce the complexity of the cavity equations for the global risk by suitable choice of the graph ensemble. While we have used this graph ensemble to study the effect of heterogeneity in transmissibility, we can also choose that the transmissibility of each group is the same, such that the groups of nodes now only vary in their mean degree and assortativity between different groups. Therefore, this ensemble provides a low complexity method to study the effect of degree correlations on epidemic risk, where correlations are driven by some hidden variable, perhaps reflecting differences in demographic.

	The distributional equations of risk give insight into the impact of different vaccination strategies and social distancing behaviours beyond their effect on the global risk. This is particularly interesting with respect to the tails of these distributions, corresponding to the probability of high risk nodes. We have shown that assortative graphs have fat tailed distributions, and a further avenue of work may be to elucidate more information about the tail of the risk distribution in graphs with strong degree correlations from the cavity method. Interestingly, the distributional equations allow one to study the risk distribution in the sub-network of unvaccinated nodes with no further technicalities.

	There are several limitations of our work. Firstly, the equations we provide only describe the steady state of the SIR model, but dynamics are an important consideration in the modelling of epidemics, for example the timing of peak infections is of particular interest. An interesting avenue of further work could be to extend our investigations to the dynamics of the SIR model using the dynamical cavity method, exploring how differences in transmission and assortativity affect the peak of infections. Additionally, our modelling approach is restricted to static networks, whereas realistically contact networks vary with time, as discussed in \cite{onaga2017concurrency,P.Peixoto2018}. Lastly, it is important to note that the SIR model does not allow for reinfection of individuals, making it unsuitable for the modelling of some infectious diseases where reinfection is a common occurrence. Despite these limitations our work shows that the cavity approach to the epidemic risk in the SIR model reveals rich behaviour governed by the heterogeneity in the transmissibility of individuals.

	\ack
	All authors would like to thank Reimer K\"uhn for useful discussions.

	\bibliographystyle{iopart-num-mod}
	\bibliography{SIRpaper2}
	
	\appendix

	\section{Risk with homogeneous transmission and node and link deletion}
	\label{app:homo}
	Here we show how to derive a closed set of equations for the average, or `global', risk that a node causes an epidemic across a network that is subject to node and link deletion, when 
	nodes are homogeneous in transmissibility. 
	We introduce a binary random variable $\sigma_{i} \in \{0,1\} ~\forall ~i$ to indicate if a node $i$ is deleted ($\sigma_{i}$=0) or not ($\sigma_{i}=1$) and $\tau_{ij} \in \{0,1\} ~\forall ~i,j$ to indicate if the link between $i$ and $j$ is deleted ($\tau_{ij}=0$) or not ($\tau_{ij}=1$). We consider the case where node and links are randomly and independently deleted with probability dependent on their degree, according to, 
	\begin{eqnarray}
		\mathrm{P}(\sigma_{i}|k_{i}) = \left(1 - x(k_{i})\right)\delta_{\sigma_{i},0} +  x(k_{i})\delta_{\sigma_{i},1}\label{app: sigma degree based prob} \\
		\mathrm{Q}(\tau_{ij}|k_{i},k_{j}) = \left(1 - y(k_{i},k_{j})\right)\delta_{\tau_{ij},0} +  y(k_{i},k_{j})\delta_{\tau_{ij},1}.\label{app: tau degree based prob}
	\end{eqnarray}
	For a graph with adjacency matrix $\bA$ the risk that node $i$ causes an epidemic is given by, 
	\begin{eqnarray}
		r_{i}(\bA) &=& \sigma_{i}\left[1 - \int_{0}^{\infty}\rmd t\, \gamma(t)\prod_{j\in \partial^\bA_{i}} (1 - \alpha(t) \tau_{ij} r_{j}^{(i)}(\bA))\right] 
		\label{eq: risk homo}\\
		r_{j}^{(i)}(\bA) &=& \sigma_{j}\left[1 - \int_{0}^{\infty}\rmd t\, \gamma(t)\prod_{\ell\in \partial^\bA_{j}\setminus
			i} \left(1 - \alpha(t)\tau_{j \ell}r_{\ell}^{(j)}(\bA)\right)\right].
		\label{eq: r hat homo}
	\end{eqnarray}
	Summing (\ref{eq: risk homo}) over $i$, dividing by $N$ and inserting unity in the form
	\begin{equation}
		1= \sum_\sigma \delta_{\sigma,\sigma_i}\sum_{k\geq 0} \delta_{k,|\partial_i^\bA|}  
		\prod_{j\in\partial_i^\bA}\int \rmd\hat{r}_j\,\delta(\hat{r}_j-r_j^{(i)}(\bA)) \sum_{\hat{\tau}_{j}}\delta_{\hat{\tau}_{j},\tau_{ij}}
	\end{equation}
	we obtain
	\begin{eqnarray}
		\hspace*{-1.2cm}    g(\bA) &=& \frac{1}{N}\sum_\sigma\sum_{i=1}^N\delta_{\sigma,\sigma_{i}}\sigma \Bigg\{1 -\sum_{k\geq0}  \delta_{k,k_i(\bA)}
		\left[\prod_{j\in\partial_i^\bA}\int \rmd\hat{r}_j\,\delta(\hat{r}_j-r_j^{(i)}(\bA))\sum_{\hat{\tau}_{j}}\delta_{\hat{\tau}_{j},\tau_{ij}}\right]\nonumber\\
		&&\times\int_{0}^{\infty}\rmd t \gamma(t)\prod_{j\in\partial_i^\bA} (1 - \alpha (t) \hat{\tau}_{j}\hat{r}_{j})\Bigg\}\nonumber\\ 
		&=& \sum_{\sigma}\sigma\Bigg[\mathrm{P}(\sigma)-\sum_{k\geq 0}\sum_{\btauh}\int\rmd \Brh\, \mathrm{W}_{c}(k,\sigma;\btauh,\Brh|\bA)\nonumber\\
		&&~~~~~~~~~~~~~~~~~~~\times\int_{0}^{\infty}\rmd t\gamma(t)\prod_{j\in\partial_i^\bA}\left (1 - \alpha (t)\hat{\tau}_{j} \hat{r}_{j}\right)\Bigg] \label{app: g(A)}
	\end{eqnarray}
	where we have defined $k_i(\bA)=|\partial^\bA_i|$ the degree of node $i$ in network $\bA$,
	$\Brh=(\hat{r}_1,\ldots,\hat{r}_k)$ the cavity fields of the neighbours of a 
	site with degree $k$, $\btauh=(\hat{\tau}_1,\ldots,\hat{\tau}_k)$ the labels that describe which links of a 
	site with degree $k$ have been deleted, $\mathrm{P}(\sigma) = N^{-1}\sum_{i}\delta_{\sigma,\sigma_{i}}$ the probability that a node has label $\sigma$, and 
	\begin{eqnarray}
		\mathrm{W}_{c}(k,\sigma;\btauh,\Brh|\bA) = \frac{1}{N}\sum_{i=1}^N\delta_{k,k_i(\bA)}\delta_{\sigma,\sigma_i}
		\prod_{j\in\partial_i^\bA} \delta(\hat{r}_j-r_j^{(i)}(\bA))\delta_{\hat{\tau}_{j},\tau_{ij}}
		\label{app:joint_k_r}
	\end{eqnarray}
	the likelihood that a site drawn at random in network $\bA$
	has degree $k$, label $\sigma$, and neighbours with cavity fields, $\Brh$, with the links to these neighbours deleted according to $\btauh$. We denote probability distributions of variables that are linked by an edge with $\mathrm{W}( ~ ;~ )$. We have also denoted distributions which depend upon the cavity fields $\Brh$ with a subscript $c$ and refer to them as `cavity distributions'. We note that
	integrating a cavity distribution $\mathrm{W}_c$ over its cavity fields leads to a normal (i.e. not cavity) distribution $\mathrm{W}$. Furthermore, we note that integrating (\ref{app:joint_k_r}) over $\btauh$ and 
	$\Brh$ leads to a distribution of single site variables, that we denote with $P$. For later reference, we will denote with $Q$ distributions of single link quantities, in accordance with (\ref{app: tau degree based prob}).
	
	Using Bayes relation, we then have $\mathrm{W}_{c}(k,\sigma;\btauh,\Brh|\bA)=\mathrm{P}(k,\sigma|\bA)\mathrm{W}_{c}(\btauh,\Brh|k,\bA)$ where we have used that, in the absence of correlations between node labels, and between node labels and link labels, the distribution $\mathrm{W}_{c}(\btauh,\Brh|k,\bA)$ in the cavity graph where node $i$ has been removed is independent of the label $\sigma$ of node $i$. 
	We proceed by writing  $\mathrm{W}_{c}(\btauh,\Brh|k,\bA)=\sum_{\mathbf{q}}\mathrm{W}_{c}(\boldsymbol{q},\btauh,\Brh|k,\bA)$ where $\mathbf{q} = (q_{1},...,q_{k})$ are the degrees of the neighbours of a node with degree $k$. 
	Using Bayes relations, we can write 
	$\mathrm{W}_{c}(\boldsymbol{q},\btauh,\Brh|k,\bA)=\mathrm{W}(\mathbf{q}|k,\bA)\mathrm{Q}(\btauh| k,\mathbf{q})\mathrm{W}_{c}(\Brh|\bA, k,\mathbf{q})$
	where we have used the conditional independence of $\Brh$ (when conditioned on the degrees $k$ and $\bq$), on $\btauh$, in the cavity graph where node $i$ and all its links have been removed. Here 
	$\mathrm{Q}(\btauh| k,\mathbf{q})$ is the joint distribution of deleting $k$ links according to $\btauh$. 
	Here $\mathrm{W}_{c}(\Brh|\bA, k,\mathbf{q})$ is the joint distribution of the cavity fields $\Brh$ given that the node removed from the cavity graph has degree $k$ with neighbours with degrees $\bq$. 
	By equation (\ref{app: tau degree based prob}) we have 
	that  
	$\mathrm{Q}(\btauh| k,\mathbf{q}) = \prod_{j=1}^{k}\mathrm{Q}(\hat{\tau}_{j}| k,q_{j})$. Furthermore, 
	in the limit $N\to\infty$, by 
	virtue of a locally tree-like assumption, $\mathrm{W}_{c}(\Brh|\bA,k,\mathbf{q})$ also factorises $\mathrm{W}_{c}(\Brh|\bA,k,\mathbf{q}) = \prod_{j=1}^{k} \mathrm{W}_{c}(\hat{r}_{j}|\bA,k,q_{j})$, for any $k>0$. For $k=0$, we can set $\mathrm{W}_c(\Brh|\bA,0,\mathbf{q})=\delta(\Brh)$ as the product in (\ref{eq: risk homo}) is empty and evaluates to one, regardless of $\Brh$. When we insert this into (\ref{app: g(A)}) we find,
	\begin{eqnarray}
		\hspace*{-2.5cm} g(\bA)\!=\!\sum_{\sigma, k\geq 0,\mathbf{q}}\!\sigma \mathrm{P}(k,\sigma|\bA)\mathrm{W}(\mathbf{q}|k,\bA) \left[1\!-\!  \int_{0}^{\infty}\rmd t\, \gamma(t) \prod_{j=1}^{k} \left(1 - \alpha (t) y(k,q_j)\hat{g}_{k,q_{j}}(\bA)\right) \right] \label{app: g(A) no pairwise}
	\end{eqnarray}
	where $\hat{g}_{k,q}(\bA)=\int \rmd\hat{r}\mathrm{W}_{c}(\hat{r}|\bA,k,q)\,\hat{r}$ is the average contribution to the cavity field of a random 
	node with degree $k$ from a random neighbour with degree $q$. 
	We now assume that the degrees of the neighbours of a random site with degree $k$ are independent when conditioned upon $k$ such that $\mathrm{W}({\bf q}|k,\bA) = \prod_{j=1}^{k} \mathrm{W}(q_{j}|k,\bA)$. While this assumption is only true for particular graph ensembles, it will act as a useful approximation for ensembles where it is not true, as it leads to a closed set of equations. Inserting this into (\ref{app: g(A) no pairwise}) yields,
	\begin{eqnarray}
		\hspace*{-2cm} g(\bA)=\sum_{ k\geq 0}\mathrm{P}(k|\bA)x(k)\left[1-  \int_{0}^{\infty}\rmd t\, \gamma(t) \prod_{j=1}^{k} \left(1 - \alpha (t) \tilde{g}_{k}(\bA)\right) \right] \label{app: g(A) with pairwise}
	\end{eqnarray}
	where
	\begin{eqnarray}
		\tilde{g}_{k}(\bA) = \sum_{q}\mathrm{W}(q|k,\bA) y(k , q) \hatg_{k,q} \label{eq:gtilde def}.
	\end{eqnarray}
	To make progress, we need an equation for $\tilde{g}_{k}(\bA)$,
	\begin{eqnarray}
		\hspace*{-1cm}    \tilde{g}_{k}(\bA) &= \sum_{q}\mathrm{W}(q|k,\bA) y(k , q) \int \rmd \hat{r} ~ \mathrm{W}_{c}(\hat{r}|k,q,\bA) \hat{r}  \\
		&=\sum_{q} \mathrm{W}(q|k,\bA)y(k,q)\int \rmd \Brh \frac{\mathrm{W}_{c}(k;\Brh,q|\bA)}{\mathrm{P}(k|\bA)\mathrm{W}(q|k,\bA)}\,\frac{1}{k} \sum_{j=1}^{k}
		\hat{r}_j \\
		&=\sum_{q} \mathrm{W}(q|k,\bA)y(k,q)\sum_{\bq}\int \rmd \Brh \frac{\mathrm{W}_{c}(k;\Brh,\bq|\bA)}{\mathrm{P}(k|\bA)\mathrm{W}(q|k,\bA)}\,\frac{1}{k} \sum_{j=1}^{k}
		\hat{r}_j \delta_{q,q_j}\\
		&= \sum_{q} \frac{y(k,q)}{k\mathrm{P}(k|\bA)}  \sum_{j=1}^{k}\sum_\bq
		\int \rmd \Brh 
		\frac{1}{N}\sum_{i}^N \delta_{k,k_i(\bA)}[\prod_{l\in\partial_i^{\bA}}\delta_{q_l,k_l(\bA)}\delta(\hat{r}_l- r_{l}^{(i)}(\bA))]
		\hat{r}_j \delta_{q,q_j}
		\nonumber\\
		&= \sum_{q} \frac{y(k,q)}{k\mathrm{P}(k|\bA)}  \sum_{j=1}^{k}
		\int \rmd \Brh 
		\frac{1}{N}\sum_{i}^N \delta_{k,k_i(\bA)}[\prod_{l\in\partial_i^{\bA}}\delta(\hat{r}_l- r_{l}^{(i)}(\bA))]
		\hat{r}_j \delta_{q,k_j}
		\nonumber\\
		&= \sum_{q} \frac{y(k,q)}{k\mathrm{P}(k|\bA)} \frac{1}{N} \sum_{j=1}^{k}
		\sum_{i}^N A_{ij}\delta_{k,k_i(\bA)}\delta_{q,k_j(\bA)}\hat{r}_j^{(i)}(\bA) 
	\end{eqnarray}
	where we have used Bayes relations and the definition 
	$$
	\mathrm{W}_{c}(k;\bq,\Brh|\bA)=\frac{1}{N}\sum_{i}^N \delta_{k,k_i(\bA)}[\prod_{j\in\partial_i^{\bA}}\delta_{q_j,k_j(\bA)}\delta(\hat{r}_j- r_{j}^{(i)}(\bA))]
	$$
	Hence, using the cavity equation (\ref{eq: r hat homo}) we find an expression for $\tilde{g}_k$ as follows, 
	\begin{eqnarray} 
		\hspace*{-2.2cm}\tilde{g}_k(\bA)&\!=\!& \sum_{q} \frac{y(k,q)}{N k\mathrm{P}(k|\bA)} \sum_{i,j=1}^N A_{ij} \delta_{k,k_i(\bA)}\delta_{q,k_j(\bA)}\sigma_{j}
		\Bigg[1\!-\! \int_{0}^{\infty}\!\rmd t \gamma(t)\!\!\prod_{\ell\in \partial^\bA_{j}\setminus i} \!\!\left(1 \!-\! \alpha(t) \tau_{j \ell}r_{\ell}^{(j)}\right)\Bigg]\nonumber\\
		\hspace*{-2.2cm}&\!=\!& \sum_{q} \frac{y(k,q)}{N k\mathrm{P}(k|\bA)} \Bigg[\prod_{\ell\in\partial_j^\bA\setminus i}\int \rmd\hat{r}'_\ell \delta(\hat{r}'_\ell-r_\ell^{(j)}) \sum_{\hat{\tau}'_{\ell}}\delta_{\hat{\tau}'_{\ell},\tau_{j\ell}}\Bigg]\nonumber\\
		\hspace*{-2.2cm}
		&&\times\sum_{\sigma}\sum_{i,j=1}^N A_{ij}\delta_{k,k_i(\bA)}\delta_{q,k_j(\bA)}\delta_{\sigma,\sigma_{j}} 
		\sigma \left[1 - \int_{0}^{\infty}\rmd t\, \gamma(t)\prod_{\ell=1}^{q-1}\left( 1 - \alpha(t)\hat{\tau}'_{\ell}\hat{r}'_{\ell}\right)\right] \nonumber  \\
		\hspace*{-2.2cm}
		&\!=\!&\sum_{q} \frac{y(k,q) \bar{k}(\bA)}{ k\mathrm{P}(k|\bA)}\int \rmd\Brh'\sum_{\btauh'}\sum_{\sigma}\mathrm{W}_{c}(k;q,\sigma;\btauh',\Brh'|\bA)\sigma\nonumber\\
		\hspace*{-2.2cm}
		&&\times\left[1-\int_{0}^{\infty}\rmd t\, \gamma(t)\prod_{\ell=1}^{q-1}\left( 1 - \alpha(t)\hat{\tau}'_{\ell}\hat{r}'_{\ell}\right)\right] \label{app: tilde g part way}
	\end{eqnarray}
	where 
	\begin{eqnarray}
		\hspace*{-2.5cm}\mathrm{W}_{c}(k;q,\sigma;\btauh',\Brh'|\bA)=\frac{\sum_{ij}A_{ij}\delta_{k,k_i(\bA)}\delta_{\sigma,\sigma_{j}}\,\delta_{q,k_j(\bA)}  [\prod_{\ell\in\partial_j^\bA\setminus i} \delta(\hat{r}'_\ell-r_\ell^{(j)}(\bA)) \delta_{\hat{\tau}_{\ell}',\tau_{j \ell}}]}{N\bar{k}(\bA)}
	\end{eqnarray}
	is the likelihood that a randomly drawn link connects a node 
	$i$ with degree $k$ to a node $j$ with degree $q$, label $\sigma$ and neighbours ($i$ excluded) with 
	cavity fields $\Brh'=(\hat{r}'_1,\ldots,
	\hat{r}'_{q-1})$ and links deleted according to $\btauh'=(\hat{\tau}'_1,\ldots,
	\hat{\tau}'_{q-1})$. We next use Bayes and the independence of $\btauh'$ and $\Brh'$ on $\sigma$ when conditioned on $q$
	to write
	$\mathrm{W}_{c}(k;q,\sigma;\btauh',\Brh'|\bA)=\mathrm{W}(k;q,\sigma|\bA)
	\mathrm{W}_{c}(\btauh',\Brh'|k,q,\bA)$ and write 
	$\mathrm{W}_{c}(\btauh',\Brh'|k,q,\bA)=\sum_{\bq'}\mathrm{W}_{c}(\btauh',\Brh',\bq'|k,q,\bA)$ where $\bq' = (q'_{1},\dots,q'_{q-1})$ are the degrees of the neighbours of a node with degree $q$, excluding the neighbour with degree $k$.
	Using again the independence of $\Brh'$ and $\btauh'$ 
	when conditioned on the degrees $q$ and $\bq'$, we have
	$\mathrm{W}_{c}(\btauh',\Brh',\bq'|k,q,\bA)= \mathrm{W}(\bq'|q,\bA)\mathrm{Q}(\btauh'|q,\bq')\mathrm{W}_{c}(\Brh'|q',\bq',\bA)$ where we also note that $\bq'$, $\Brh'$ and $\btauh'$ 
	are independent of $k$ when conditioned on $q$.
	By equation (\ref{app: tau degree based prob}) the $\btauh'$ factorise, in addition to the $\Brh'$ which factorise under the tree-like assumption, and we again assume that $\bq'$ factorise when conditioned on $q$, hence we have 
	\begin{equation}
		\mathrm{W}_{c}(\bq'|q,\bA)\mathrm{Q}(\btauh'|q,\bq')\mathrm{W}_{c}(\Brh'|q,\bq',\bA) = \prod_{\ell=1}^{q-1}\mathrm{W}(q'_{\ell}|q,\bA)\mathrm{Q}(\hat{\tau}'_{\ell}|q,q'_{\ell})\mathrm{W}_{c}(\hat{r}'_{\ell}|q,q'_{\ell},\bA).
	\end{equation}
	Inserting this into (\ref{app: tilde g part way}) leads to the system of equations,
	\begin{eqnarray}
		\hspace{-1cm}\tilde{g}_k(\bA)&=&\sum_{q\geq 1} \sum_{\sigma} y(k,q) \frac{\mathrm{W}(k;q,\sigma|\bA)}{\mathrm{W}(k|\bA)}\sigma\nonumber\\
		&&\hspace{-0.5cm}\times\left[1- \int_{0}^{\infty}\rmd t\, \gamma(t)\left( 1 - \alpha(t) \sum_{q'} \mathrm{W}(q'|q,\bA) y(q,q') \hat{g}_{q,q'}(\bA)\right)^{q-1}\right] 
	\end{eqnarray}
	where $\mathrm{W}(k|\bA)=\sum_{q}\mathrm{W}(k;q|\bA)=k\mathrm{P}(k|\bA)/\bar{k}(\bA)$ and
	the sum over $q$ has been restricted to $q\geq 1$, as 
	$\mathrm{W}(k;q|\bA)$ vanishes for $q=0$.
	Finally, by 
	using (\ref{eq:gtilde def})
	we have the following set of equations 
	\begin{eqnarray}
		\hspace*{-1cm}g(\bA)&=&\sum_{k\geq 0}\mathrm{P}(k|\bA)x(k)
		\left[1\!-\!  \int_{0}^{\infty}\!\!\!\!\rmd t\, \gamma(t)\! \left(1 - \alpha (t) \tilde{g}_k(\bA)\right)^{k} \right]
		\label{eq: cavity risk hetero global} \\
		\hspace*{-1cm}\tilde{g}_k(\bA)& =& \sum_{q\geq 1}  y(k,q)\mathrm{W}(q|k,\bA) x(q)
		\left[1 \!-\! \int_{0}^{\infty} \rmd t\, \gamma(t)\left( 1 - \alpha(t) \tilde{g}_{q}(\bA)\right)^{q-1} \right]
		\label{eq: cavity risk hetero iterative}
	\end{eqnarray}
	where we have used (\ref{app: sigma degree based prob}) and
	denoted $\mathrm{W}(q|k,\bA)=\mathrm{W}(k;q|\bA)/\mathrm{W}(k|\bA)$.
	These show that the global risk $g(\bA)$ has the same 
	self-averaging properties of $\mathrm{P}(k|\bA)$ and $\mathrm{W}(k;q|\bA)$. 
	If the set of equations (\ref{eq: cavity risk hetero iterative}) is solved numerically, the solution can be substituted into equation (\ref{eq: cavity risk hetero global}) to find the global risk.
	
	\section{Risk with heterogeneous transmission and link deletion}\label{app:hetero}
	In this section we find closed equations for the global risk when nodes have heterogeneous transmissibility. For brevity we do not consider node deletion in this section. We consider link deletion but now based upon the label and degree of nodes, such that, 
	\begin{eqnarray}
		\mathrm{Q}(\tau_{ij}) = \left(1- y(k_{i},\xi_{i},k_{j},\xi_{j}) \right)\delta_{\tau_{ij},0} + y(k_{i},\xi_{i},k_{j},\xi_{j})\delta_{\tau_{ij},1} \label{app:hetero label tau dist}
	\end{eqnarray}
	where $1 - y(k,\xi,k',\xi')$ is the probability of deleting a link with a node with degree $k$ and label $\xi$ at one end and a node with 
	degree $k'$ and label $\xi'$ at the other. 
	Summing (\ref{eq: r_i hetero link percolation}) over $i$, dividing by $N$ and inserting unity in the form
	\begin{equation}
		1= \sum_\sigma \delta_{\sigma,\sigma_i}\sum_{k\geq 0} \delta_{k,|\partial_i^\bA|}  
		\prod_{j\in\partial_i^\bA}\int \rmd\hat{r}_j\,\delta(\hat{r}_j-r_j^{(i)}(\bA)) \sum_{\hat{\tau}_{j}}\delta_{\hat{\tau}_{j},\tau_{ij}}
	\end{equation}
	we obtain
	\begin{eqnarray}
		\hspace*{-2cm}g(\bA,\bxi) &=& 1-\frac{1}{N}\sum_{i=1}^N\sum_{k\geq0}  \delta_{k,k_i(\bA)}\sum_\xi \delta_{\xi,\xi_i} \left[\prod_{j\in\partial_i^\bA}\int \rmd\hat{r}_j\,\delta(\hat{r}_j-r_j^{(i)}(\bA,\bxi))\sum_{\hat{\tau}_{j}}\delta_{\hat{\tau}_{j},\tau_{ij}}\right]
		\nonumber\\
		&&\times
		\int_{0}^{\infty}\rmd t \gamma(t|\xi)\prod_{j\in\partial_i^\bA} (1 - \alpha (t) \hat{\tau}_{j}\hat{r}_{j}) \nonumber\\ 
		&=& 1-\sum_{k\geq 0,\xi,\btauh}\int\rmd \Brh\, \mathrm{W}_{c}(k,\xi;\Brh,\btauh|\bA,\bxi)\int_{0}^{\infty}\rmd t\, \gamma(t|\xi)\prod_{j\in\partial_i^\bA}\left (1 - \alpha (t) \hat{\tau}_{j}  \hat{r}_{j}\right)\nonumber\\
		\label{eq:hetero-risk}
	\end{eqnarray}
	where we have defined  
	$\Brh=(\hat{r}_1,\ldots,\hat{r}_k)$ and 
	\begin{eqnarray}
		\mathrm{W}_{c}(k,\xi;\Brh,\btauh|\bA,\bxi) = \frac{1}{N}\sum_{i=1}^N\delta_{k,k_i(\bA)} \delta_{\xi,\xi_i}
		\prod_{j\in\partial_i^\bA} \delta(\hat{r}_j-r_j^{(i)}(\bA,\bxi))\delta_{\hat{\tau}_{j},\tau_{ij}}
		\label{eq:Pk-xi-r}
	\end{eqnarray}
	the likelihood that a site drawn at random in network $\bA$
	has degree $k$, label $\xi$ and neighbours with cavity fields $\Brh=(\hat{r}_1,\ldots,
	\hat{r}_{k})$, and links deleted according to $\btauh=(\hat{\tau}_1,\ldots,
	\hat{\tau}_{k})$. 
	Similarly to before, we use 
	Bayes relation to write
	$\mathrm{W}_{c}(k,\xi;\Brh,\btauh|\bA,\bxi)=\mathrm{P}(k,\xi|\bA,\bxi)\mathrm{W}_{c}(\Brh,\btauh|k,\xi,\bA,\bxi)$ 
	and we then write
	$\mathrm{W}_{c}(\Brh,\btauh|k,\xi,\bA,\bxi) = \sum_{\bzet,\bq}\mathrm{W}_{c}(\Brh,\btauh,\bzet,\bq|k,\xi,\bA,\bxi)$, where $\bzet=(\zeta_1,\ldots,
	\zeta_{k})$ and $\bq=(q_1,\ldots,
	q_{k})$ are the labels and degrees of the neighbours of a random site with degree $k$ and label $\xi$. Using Bayes theorem this becomes $\mathrm{W}_{c}(\Brh,\btauh,\bzet,\bq|k,\xi,\bA,\bxi) =\mathrm{W}(\bzet,\bq|k,\xi,\bA,\bxi)\mathrm{Q}(\btauh|k,\xi,\bzet,\bq)\mathrm{W}_{c}(\Brh|k,\xi,\bzet,\bq,\bA,\bxi) $ where we have also used the independence of $\btauh$ and $\Brh$ when conditioned on $\bzet$ and $\bq$. For $k=0$ we can set $\mathrm{W}_{c}(\Brh|0,\xi,\bzet,\bq,\bA,\bxi) =\delta(\Brh)$ as the products in (\ref{eq:hetero-risk}) evaluate to one. In the limit $N\to\infty$, by 
	virtue of a locally tree-like assumption, the joint distribution of the cavity fields $\Brh$ factorises 
	$\mathrm{W}_{c}(\Brh|k,\xi,\bzet,\bq,\bA,\bxi)=\prod_{j=1}^{k}\mathrm{W}_{c}(\hat{r}_{j}|k,\xi,q_{j},\zeta_{j},\bA,\bxi)$, as does $\mathrm{Q}(\btauh|k,\xi,\bzet,\bq)=\prod_{j=1}^{k}\mathrm{Q}(\hat{\tau}_{j}|k,\xi,q_{j},\zeta_{j})$ by equation (\ref{app:hetero label tau dist}). Again, we assume the degrees and labels of the neighbours of a random site with degree $k$ and label $\xi$, are independent, when conditioned on the degree $k$ and label $\xi$, such that $\mathrm{W}(\bzet,\bq|k,\xi,\bA,\bxi) = \prod_{j=1}^{k} \mathrm{W}(\zeta_{j},q_{j}|k,\xi,\bA,\bxi)$. As noted above, although this is only true for particular graph ensembles, it acts as a useful approximation for ensembles where it is not true, as it leads to a closed set of equations. When this is inserted into (\ref{eq:hetero-risk}) we find,
	\begin{eqnarray}
		g(\bA,\bxi)=1- \sum_{k,\xi}\mathrm{P}(k,\xi|\bA,\bxi) \int_{0}^{\infty}\rmd t\, \gamma(t|\xi) \left(1 - \alpha (t) \tilde{g}_{k, \xi}(\bA,\bxi)\right)^{k} 
		\label{eq:g-hetero-A}
	\end{eqnarray}
	where $\tilde{g}_{k, \xi}(\bA,\bxi)=\sum_{q,\xi'}\mathrm{W}(q,\zeta|k,\xi,\bA,\bxi)y(k,\xi,q,\zeta)\int \rmd\hat{r}\mathrm{W}_{c}(\hat{r}'|k,\xi,q,\zeta,\bA,\bxi)\,\hat{r}$ is the average cavity field of a random neighbour of a random 
	node with degree $k$ and label $\xi$. To make progress, we need an equation for $\tilde{g}_{k, \xi}(\bA,\bxi)$.
	To this end we can write 
	$\tilde{g}_{k ,\xi}(\bA,\bxi)$ as 
	\begin{eqnarray}
		\hspace*{-2cm}\hat{g}_{k, \xi}(\bA,\bxi)&=& \sum_{q,\zeta}\mathrm{W}(q,\zeta|k,\xi,\bA,\bxi)y(k,\xi,q,\zeta)\int \rmd\hat{r}\mathrm{W}_{c}(\hat{r}|k,\xi,q,\zeta,\bA,\bxi)\,\hat{r}
		\nonumber\\
		&=&\sum_{q,\zeta}y(k,\xi,q,\zeta)\int \rmd \Brh\, \frac{\mathrm{W}_{c}(k,\xi;q,\zeta,\Brh|\bA,\bxi)}{\mathrm{P}(k,\xi|\bA,\bxi)}\,\frac{1}{k} \sum_{j=1}^{k}  \hat{r}_j 
		\nonumber\\
		&=& \sum_{q,\zeta}y(k,\xi,q,\zeta)\sum_{\bq,\bzeta}\int \rmd \Brh\, \frac{\mathrm{W}_{c}(k,\xi;\bq,\bzeta,\Brh|\bA,\bxi)}{\mathrm{P}(k,\xi|\bA,\bxi)}\,\frac{1}{k} \sum_{j=1}^{k}
		\hat{r}_j \delta_{q,q_{j}}\delta_{\zeta,\zeta_{j}}
		\nonumber\\
		&=&\sum_{q,\zeta}\frac{y(k,\xi,q,\zeta)}{k \mathrm{P}(k,\xi|\bA,\bxi)} \sum_{j=1}^{k}\sum_{\bq,\bzeta} \int \rmd \Brh\, \frac{1}{N}\sum_{i}\delta_{k,k_{i}(\bA)}\delta_{\xi,\xi_{i}}\nonumber \\
		&&\times\left[\prod_{l\in\partial_{i}^{\bA}}\delta(\hat{r}_{l}-r_{l}^{(i)}(\bA,\bxi))\delta_{q_{l},k_{l}(\bA)}\delta_{\zeta_{l},\xi_{l}}\right]\, 
		\hat{r}_j  \delta_{q,q_{j}}\delta_{\zeta,\zeta_{j}}
		\nonumber\\
		&=& \sum_{q,\zeta}\frac{y(k,\xi,q,\zeta)}{N k \mathrm{P}(k,\xi|\bA,\bxi)}\sum_{i,j=1}^{N}A_{ij}\delta_{k,k_{i}(\bA)}\delta_{\xi,\xi_{i}}\delta_{q,k_{j}(\bA)}\delta_{\zeta,\xi_{j}} r_{j}^{(i)}(\bA,\bxi)
	\end{eqnarray}
	Hence, using the cavity equation (\ref{eq: rj_i hetero link percolation}) we find an expression for $\tilde{g}_{k,\xi}$ as follows, 
	\begin{eqnarray}
		\tilde{g}_{k, \xi}(\bA,\bxi)&=&\sum_{q,\zeta}\frac{y(k,\xi,q,\zeta)}{N k \mathrm{P}(k,\xi|\bA,\bxi)}\sum_{i,j=1}^{N}A_{ij}\delta_{k,k_{i}(\bA)}\delta_{\xi,\xi_{i}}\delta_{q,k_{j}(\bA)}\delta_{\zeta,\xi_{j}}r_{j}^{(i)}(\bA,\bxi) \nonumber\\
		&=&\sum_{q,\zeta}\frac{y(k,\xi,q,\zeta)}{N k \mathrm{P}(k,\xi|\bA,\bxi)}\sum_{i,j=1}^{N}A_{ij}\delta_{k,k_{i}(\bA)}\delta_{\xi,\xi_{i}}\delta_{q,k_{j}(\bA)}\delta_{\zeta,\xi_{j}} \nonumber \\ &&\times\left\{\prod_{\ell\in\partial_j^\bA\setminus i}\int \rmd\hat{r}'_\ell\, \delta(\hat{r}'_\ell-r_\ell^{(j)}(\bA,\bxi)) \sum_{\hat{\tau}'_{\ell}}\delta_{\hat{\tau}'_{\ell},\tau_{j \ell}}\right\}\nonumber\\
		&&\times\left[1 -  \int_{0}^{\infty}\rmd t\, \gamma(t|\zeta)\prod_{\ell\in \partial^\bA_{j}\setminus i} \left(1 - \alpha(t)\hat{\tau}'_{\ell}\hat{r}'_{\ell}\right)\right]\nonumber\\
		&=&\sum_{q,\zeta}\frac{y(k,\xi,q,\zeta)}{ k \mathrm{P}(k,\xi|\bA,\bxi)} \sum_{\btauh'}\int \rmd \Brh' \mathrm{W}_{c}(k,\xi ;q,\zeta;\btauh',\Brh'|\bA,\bxi) \nonumber\\
		&&\times\left[1 -  \int_{0}^{\infty}\rmd t\, \gamma(t|\zeta)\prod_{\ell=1}^{q-1} \left(1 - \alpha(t)\hat{\tau}'_{\ell}\hat{r}'_{\ell}\right)\right]\nonumber\\\label{app:hetero-g-tau-near}
	\end{eqnarray}
	where 
	\begin{eqnarray}
		\hspace*{-2cm}\mathrm{W}_{c}(k,\xi;q,\zeta;\btauh',\Brh'|\bA,\bxi)=\frac{\sum_{ij}A_{ij}\delta_{k,k_i(\bA)}\,\delta_{q,k_j(\bA)}\delta_{\xi,\xi_i} \delta_{\zeta,\xi_j} \prod_{\ell\in\partial_j^\bA\setminus i} \delta(\hat{r}'_\ell-r_\ell^{(j)}(\bA,\bxi))\delta_{\hat{\tau}'_{\ell},\tau_{j \ell}}}{N\bar{k}(\bA)} \nonumber \\
	\end{eqnarray}
	is the likelihood that a randomly drawn link connects a node 
	$i$ with degree $k$ and label $\xi$ to a node $j$ with degree $q$, label $\zeta$ and neighbours (excluded $i$) with 
	cavity fields $\Brh'=(\hat{r}'_1,\ldots,
	\hat{r}'_{q-1})$ with links deleted according to $\btauh' = ( \hat{\tau}'_{1},\ldots,\hat{\tau}'_{q-1})$. By Bayes theorem 
	$\mathrm{W}_{c}(k,\xi;q,\zeta;\btauh',\Brh'|\bA,\bxi) = \mathrm{W}(k,\xi;q,\zeta|\bA,\bxi)\mathrm{W}_{c}(\btauh',\Brh'|k,\xi,q,\zeta,\bA,\bxi)$
	and we my 
	then write $\mathrm{W}_{c}(\btauh',\Brh'|k,\xi,q,\zeta,\bA,\bxi) = \sum_{\bq',\bzet'}\mathrm{W}_{c}(\bq',\bzet',\btauh',\Brh'|k,\xi,q,\zeta,\bA,\bxi)$ where $\bq' = (q'_{1},\ldots,q'_{q-1})$ and $\bzet' = (\zeta'_{1},\ldots,\zeta'_{q-1})$ are the degrees and labels of the neighbours of a site with degree $q$ and label $\zeta$ which is itself connected to a site with degree $k$ and label $\xi$. By Bayes theorem we may write, $\mathrm{W}_{c}(\bq',\bzet',\btauh',\Brh'|k,\xi,q,\zeta,\bA,\bxi) = \mathrm{W}(\bq',\bzet'|q,\xi,\bA,\bxi)\mathrm{Q}(\btauh',\Brh'|q,\xi,\bq',\bzet')\mathrm{W}_{c}(\Brh'|q,\xi,\bq',\bzet',\bA,\bxi)$, and noted that $\btauh'$ and $\Brh'$ are independent when conditioned on the labels and degrees. We have also noted that $\bq'$, $\bzet'$, $\btauh'$ and $\Brh'$ are independent of $k$ and $\xi$. By equation (\ref{app:hetero label tau dist}) we have $\mathrm{Q}(\btauh'|q,\xi,\bq',\bzet')=\prod_{l=1}^{q-1}\mathrm{Q}(\hat{\tau}'_{\ell}|q,\xi,q'_{\ell},\zeta'_{\ell})$ and by the tree-like assumption $\mathrm{W}_{c}(\Brh'|q,\xi,\bq',\bzet',\bA,\bxi) =\prod_{\ell=1}^{q-1} \mathrm{W}_{c}(\hat{r}'_{\ell}|q,\xi,q'_{\ell},\zeta'_{\ell},\bA,\bxi)$. Furthermore, we again assume that degrees and labels of neighbours factorise when conditioned upon the degree and labels of their neighbours i.e. $\mathrm{W}(\bq',\bzet'|q,\xi,\bA,\bxi) = \prod_{\ell=1}^{q-1}\mathrm{W}(q'_{\ell},\zeta'_{\ell}|q,\xi,\bA,\bxi)$. Inserting this into (\ref{app:hetero-g-tau-near}) leads to a closed set of equations for $\tilde{g}_{k ,\xi}(\bA,\bxi)$,
	\begin{eqnarray}
		\hspace*{-1cm}\tilde{g}_{k, \xi}(\bA,\bxi)&=&\sum_{q,\zeta}y(k,\xi,q,\zeta)\mathrm{W}(q,\zeta|k,\xi,\bA,\bxi) \nonumber\\
		&&\times \left[1 - \int_{0}^{\infty}\rmd t\, \gamma(t|\zeta)\left( 1 - \alpha(t)\,\tilde{g}_{q, \zeta}(\bA,\bxi)\right)^{q-1} \right] 
		\label{eq:ghat-hetero-A}
	\end{eqnarray}
	Assuming that $\mathrm{P}(k,\xi|\bA,\bxi)$ and $\mathrm{W}(q,\zeta|k,\xi,\bA,\bxi)$ are self-averaging over the graph
	ensemble and the node label distribution, we finally
	average equations (\ref{eq:g-hetero-A}) and (\ref{eq:ghat-hetero-A}) over the distribution $\mathrm{P}(\bA,\bxi)$ of graphs and node labels, obtaining
	\begin{eqnarray}
		\hspace*{-1cm} g &= 1 - \sum_{k,\xi}\mathrm{P}(k,\xi)\int_{0}^{\infty} \rmd t\, \gamma(t|\xi)\left(1 - \alpha(t)\,\tilde{g}_{k,\xi}\right)^{k} \label{eq:g-hetero} \\
		\hspace*{-2cm}\tilde{g}_{k, \xi}&=\sum_{q,\zeta}y(k,\xi,q,\zeta)\mathrm{W}(q,\zeta|k,\xi) \left[1 - \int_{0}^{\infty}\rmd t\, \gamma(t|\zeta)\left( 1 - \alpha(t)\,\tilde{g}_{q, \zeta}\right)^{q-1} \right]  \label{eq:ghat-hetero}
	\end{eqnarray}
	where $\mathrm{P}(k,\xi)=\bra \mathrm{P}(k,\xi|\bA,\bxi)\ket_{\bA,\bxi}$ and $\mathrm{W}(q,\zeta|k,\xi)=\bra \mathrm{W}(q,\zeta|k,\xi,\bA,\bxi)\ket_{\bA,\bxi}$ with $\bra \cdot\ket_{\bA,\bxi}=\sum_{\bA,\bxi} \cdot \mathrm{P}(\bA,\bxi)$.

	\section{An ensemble of networks linking nodes of similar or dissimilar transmissability}\label{sec: app degree dist}

	In this section we derive relations for the average degree distribution $\mathrm{P}(k|\bxi)$ and average degree correlations
	$\mathrm{W}(k,\xi;k',\xi'|\bxi)$ for networks drawn from the ensemble
	\begin{eqnarray}
		\mathrm{P}(\boldsymbol{A}| \boldsymbol{\xi}) &= \prod_{i<j} \left[\frac{\bra k \ket}{N} \frac{\mathrm{W}(\xi_{i};\xi_{j})}{\mathrm{P}(\xi_{i})\mathrm{P}(\xi_{j})}\delta_{A_{ij},1} + \left(1 - \frac{\bra k \ket}{N} \frac{\mathrm{W}(\xi_{i};\xi_{j})}{\mathrm{P}(\xi_{i})\mathrm{P}(\xi_{j})}\right)\delta_{A_{ij},0} \right] \label{app: p A given xi}
	\end{eqnarray}
	considered in (\ref{eq: ModER ensemble}), in the limit of large network 
	size $N\to \infty$.
	We start by computing the ensemble average 
	\begin{eqnarray}
		\mathrm{P}(k|\boldsymbol{\xi}) = \sum_{\boldsymbol{A}}\mathrm{P}(\boldsymbol{A}|\boldsymbol{\xi})\mathrm{P}(k|\boldsymbol{A}) \label{app: p k given xi}
	\end{eqnarray} 
	of the degree distribution for the single network instance
	\begin{eqnarray}
		\label{app: p k given A}
		\mathrm{P}( k | \boldsymbol{A},\bxi) &= \frac{1}{N}\sum_{i=1}^N\delta_{k,\sum_{j(\neq i)}A_{ij}} \end{eqnarray}
	Inserting (\ref{app: p k given A}) and (\ref{app: p A given xi}) into (\ref{app: p k given xi}), writing $\mathrm{P}(\bA|\bxi)=\prod_{i<j}\mathrm{P}(A_{ij}|\bxi)$ and using Fourier representations of Kronecker delta-functions, as well as the symmetric property of the adjacency matrix $A_{ij}=A_{ji}$,
	we get
	\begin{eqnarray}
		\mathrm{P}(k | \boldsymbol{\xi}) &=  \frac{1}{N} \sum_{i} \sum_{\boldsymbol{A}} \int \frac{\rmd\omega}{2\pi}\rme^{\rmi\omega k - \rmi \omega \sum_{j(\neq i)}A_{ij}}\prod_{k<j} \mathrm{P}(A_{kj}|\bxi)\nonumber \\
		&=  \frac{1}{N} \sum_{i} \sum_{\boldsymbol{A}} \int \frac{\rmd\omega}{2\pi}\rme^{\rmi\omega k - \rmi \omega \sum_{k\neq j}A_{kj}\delta_{ki}}\prod_{k<j} \mathrm{P}(A_{kj}|\bxi) \nonumber\\
		&=  \frac{1}{N} \sum_{i} \sum_{\boldsymbol{A}} \int \frac{\rmd\omega}{2\pi}\rme^{i\omega k - i \omega \sum_{k<j}A_{kj}(\delta_{ki}+\delta_{ji})}\prod_{k<j} \mathrm{P}(A_{kj}|\bxi)\nonumber \\
		&=  \frac{1}{N} \sum_{i} \int \frac{d\omega}{2\pi}\rme^{i\omega k}\prod_{k<j}\sum_{A_{kj}}\mathrm{P}(A_{kj}|\bxi)\, \rme^{ - i \omega A_{kj}(\delta_{ki}+\delta_{ji})} .
	\end{eqnarray}
	Taking the sum over $A_{kj}$
	and employing the (asymptotic) identity $1 + x/N = \rme^{x/N +O(1/N^2)}$, we have
	\begin{eqnarray}
		\mathrm{P}(k | \boldsymbol{\xi})&=  \frac{1}{N} \sum_{i} \int \frac{\rmd\omega}{2\pi}\rme^{\rmi\omega k}\prod_{k<j} \rme^{ \frac{\bra k \ket}{N} \frac{\mathrm{W}(\xi_{k};\xi_{j})}{\mathrm{P}(\xi_{k})\mathrm{P}(\xi_{j})} \left( \rme^{ - \rmi \omega(\delta_{ki}+\delta_{ji})} - 1 \right)}\nonumber \\
		&=  \frac{1}{N} \sum_{i} \int \frac{\rmd\omega}{2\pi}\rme^{\rmi\omega k}\prod_{k<j} \rme^{ \frac{\bra k \ket}{N} \frac{\mathrm{W}(\xi_{k};\xi_{j})}{\mathrm{P}(\xi_{k})\mathrm{P}(\xi_{j})} \left( (\rme^{ - \rmi \omega} - 1)(\delta_{ki}+\delta_{ji}) \right)}\nonumber \\
		&=  \frac{1}{N} \sum_{i} \int \frac{\rmd\omega}{2\pi}\rme^{\rmi\omega k}\rme^{ \frac{\bra k \ket}{2N}\sum_{k\neq j}  \frac{\mathrm{W}(\xi_{k};\xi_{j})}{\mathrm{P}(\xi_{k})\mathrm{P}(\xi_{j})} \left( (\rme^{ - \rmi \omega} - 1)(\delta_{ki}+\delta_{ji}) \right)}\nonumber \\
		&=  \frac{1}{N} \sum_{i} \int \frac{\rmd\omega}{2\pi}\rme^{\rmi\omega k}\rme^{ (\rme^{ - \rmi \omega} - 1)\frac{\bra k \ket}{N} \sum_{k (\neq i)}  \frac{\mathrm{W}(\xi_{k};\xi_{i})}{\mathrm{P}(\xi_{k})\mathrm{P}(\xi_{i})}}. 
	\end{eqnarray}
	At this point we make use of properties of the kernel $\mathrm{W}(\xi;\xi')$ as follows, 
	\begin{eqnarray}
		\frac{\bra k \ket}{N} \sum_{k (\neq i)}  \frac{\mathrm{W}(\xi_{k};\xi_{i})}{\mathrm{P}(\xi_{k})\mathrm{P}(\xi_{i})}  
		&= \frac{\bra k \ket}{N} \sum_{k (\neq i)} \sum_{\xi}\delta_{\xi,\xi_{k}} \frac{\mathrm{W}(\xi;\xi_{i})}{\mathrm{P}(\xi)\mathrm{P}(\xi_{i})}\nonumber  \\
		&= \bra k \ket \sum_{\xi}\mathrm{P}(\xi) \frac{\mathrm{W}(\xi;\xi_{i})}{\mathrm{P}(\xi)\mathrm{P}(\xi_{i})}=\bra k \ket\frac{\mathrm{W}(\xi_{i})}{\mathrm{P}(\xi_{i})}
	\end{eqnarray}
	where we have used the law of large numbers $\mathrm{P}(\xi)=\lim_{N\to\infty}\frac{1}{N} \sum_{i=1}^N\delta_{\xi,\xi_{i}}$ and neglected ${\mathcal O}(N^{-1})$ terms.  This allows us to derive the following result
	\begin{eqnarray}
		\mathrm{P}(k|\boldsymbol{\xi})&=  \frac{1}{N} \sum_{i} \int \frac{\rmd\omega}{2\pi}\rme^{\rmi\omega k}\rme^{ (\rme^{ - \rmi \omega} - 1)\bra k \ket\frac{\mathrm{W}(\xi_{i})}{\mathrm{P}(\xi_{i})}}\nonumber \\
		&=  \frac{1}{N} \sum_{i} \sum_{\xi}\delta_{\xi,\xi_{i}} \int \frac{\rmd\omega}{2\pi}\rme^{\rmi\omega k}\rme^{ (\rme^{ - \rmi \omega} - 1)\bra k \ket\frac{\mathrm{W}(\xi)}{\mathrm{P}(\xi)}} \nonumber\\
		&=  \sum_{\xi}\mathrm{P}(\xi)\,\rme^{ -\bra k \ket\frac{\mathrm{W}(\xi)}{\mathrm{P}(\xi)}} \int \frac{\rmd\omega}{2\pi}\rme^{\rmi\omega k}\,\rme^{\rme^{ - \rmi \omega}\bra k \ket\frac{\mathrm{W}(\xi)}{\mathrm{P}(\xi)}}\nonumber \\
		&=  \sum_{\xi}\mathrm{P}(\xi)\,\rme^{ -\bra k \ket\frac{\mathrm{W}(\xi)}{\mathrm{P}(\xi)}} \int \frac{\rmd\omega}{2\pi}\rme^{\rmi\omega k} \sum_{\ell \geq 0} \frac{(\bra k \ket\frac{\mathrm{W}(\xi)}{\mathrm{P}(\xi)})^{\ell}}{\ell!}\rme^{ - \rmi \omega \ell}\nonumber \\
		&=  \sum_{\xi}\mathrm{P}(\xi)\,\rme^{ -\bra k \ket\frac{\mathrm{W}(\xi)}{\mathrm{P}(\xi)}} \frac{(\bra k \ket\frac{\mathrm{W}(\xi)}{\mathrm{P}(\xi)})^{k}}{k!}.
	\end{eqnarray}
	The above can be simplified further. Starting from the definition of the kernel $ \mathrm{W}(\xi;\xi') = \frac{\sum_{i \neq j } A_{ij} \delta_{\xi,\xi_{i}}\delta_{\xi',\xi_{j}}}{\sum_{i\neq j}A_{ij}}$ we can write an expression for the marginal distribution $\mathrm{W}(\xi)$,
	\begin{eqnarray}
		\mathrm{W}(\xi) &= \sum_{\xi'}  \frac{\sum_{i \neq j } A_{ij} \delta_{\xi,\xi_{i}}\delta_{\xi',\xi_{j}}}{\sum_{i\neq j}A_{ij}} =  \frac{\sum_{i \neq j } A_{ij} \delta_{\xi,\xi_{i}}}{\sum_{i\neq j}A_{ij}}=\frac{\sum_i k_i \delta_{\xi,\xi_i}}{N\bra k \ket}\nonumber \\
		&= \frac{1}{N \bra k \ket} \sum_{k}\sum_{i} k \delta_{\xi,\xi_{i}} \delta_{k,k_{i}} = \frac{1}{\bra k \ket}\sum_{k} k \mathrm{P}(\xi,k) \nonumber\\
		&= \frac{1}{\bra k \ket}\mathrm{P}(\xi)\sum_{k} k \mathrm{P}(k|\xi) = \frac{\bar{k}(\xi)}{\bra k \ket} \mathrm{P}(\xi)
	\end{eqnarray}
	where we have defined $\bar{k}(\xi) = \sum_{k}\mathrm{P}(k|\xi)k$. We can now substitute this back into our previous expression to find, 
	\begin{eqnarray}
		\mathrm{P}(k | \boldsymbol{\xi}) 
		&=  \sum_{\xi}\mathrm{P}(\xi)\rme^{ -\bar{k}(\xi)} \frac{(\bar{k}(\xi))^{k}}{k!}.\label{app: degree dist}
	\end{eqnarray}
	We note that the dependence on the specific realisation of $\boldsymbol{\xi}$ is lost and only the distribution $\mathrm{P}(\xi)$ enters the expression, hence we will write $\mathrm{P}(k|\boldsymbol{\xi}) = \mathrm{P}(k)$.
	We note that by writing the degree distribution as, 
	\begin{eqnarray}
		\mathrm{P}(k) = \sum_{\xi} \mathrm{P}(\xi) \mathrm{P}(k|\xi)
	\end{eqnarray}
	comparison with (\ref{app: degree dist}) reveals the degree distribution of nodes with the same label is the Poissonian distribution
	\begin{eqnarray}
		\mathrm{P}(k|\xi) = \rme^{ -\bar{k}(\xi)} \frac{(\bar{k}(\xi))^{k}}{k!},
	\end{eqnarray}
	with mean degree $\bar{k}(\xi)$. We also note that the joint distribution $\mathrm{P}(k,\xi)$, which is required to solve 
	(\ref{eq:hetero-global-risk}), is obtained multiplying the above times $\mathrm{P}(\xi)$
	\begin{eqnarray}
		\mathrm{P}(k,\xi) = \mathrm{P}(\xi)\rme^{ -\bar{k}(\xi)} \frac{(\bar{k}(\xi))^{k}}{k!}.
	\end{eqnarray}
	We can follow a similar procedure to find the average 
	value of degree correlations in the ensemble (\ref{app: p A given xi})
	\begin{eqnarray}
		\mathrm{W}(k,\xi;k',\xi' | \bxi) &= \sum_{\bA} \mathrm{P}(\bA|\bxi) \frac{1}{N \bra k \ket} \sum_{i j} A_{ij} \delta_{k,k_{i}(\bA)} \delta_{k',k_{j}(\bA)} \delta_{\xi,\xi_{i}}\delta_{\xi',\xi_{j}}.
	\end{eqnarray}
	By using Fourier representations of the Kronecker delta-function and taking the sum over $\bA$ we find, in the large $N$ limit, 
	\begin{eqnarray}
		\mathrm{W}(k,\xi;k',\xi' | \bxi) &= \mathrm{W}(\xi;\xi')\mathrm{W}(k|\xi)\mathrm{W}(k'|\xi')
	\end{eqnarray}
	with $\mathrm{W}(k|\xi)=k\mathrm{P}(k|\xi)/\bar{k}(\xi)$. This again reveals that the degree correlations, when averaged over the graph ensemble, no longer depend on the specific configuration $\bxi$, so we will write $\mathrm{W}(k,\xi;k',\xi'|\bxi) =\mathrm{W}(k,\xi;k',\xi')$. 
	
\end{document}